%% file: main_final.tex
\documentclass[fleqn,usenatbib]{mnras}

\usepackage{newtxtext,newtxmath}

\usepackage[T1]{fontenc}
\usepackage{ae,aecompl}
\usepackage[normalem]{ulem}

\usepackage{graphicx}	
\usepackage{amsmath}	
\usepackage{multirow}
\usepackage{footnote}
\usepackage{multicol}
\usepackage{booktabs}
\usepackage{import}
\usepackage{hyperref}
\usepackage[export]{adjustbox}
\usepackage{subcaption}
\usepackage{caption}
\usepackage[normalem]{ulem}

\newcommand*\black{\color{black}}

\title[Simultaneous estimation of phasing errors and NCPA]{Simultaneous estimation of segmented telescope phasing errors and non-common path aberrations from adaptive optics corrected images}

\author[M. P. Lamb et al.]{Masen P. Lamb,$^{1}$\thanks{E-mail: masen.lamb@dunlap.utoronto.ca}
Carlos Correia,$^{2}$
Suresh Sivanandam,$^{1,3}$
Robin Swanson$^{1,4}$
\newauthor
and Polina Zavyalova$^{3,5}$
\\
$^{1}$Dunlap Institute for Astronomy and Astrophysics, University of Toronto, 50 St George Street, Toronto ON, M5S 3H4, Canada\\
$^{2}$Space ODT, Rua A. C. Monteiro, 65, 4050-014, Porto, Portugal\\
$^{3}$David A. Dunlap Department of Astronomy and Astrophysics, University of Toronto, 50 St George Street, Toronto ON, M5S 3H4, Canada\\
$^{4}$Department of Computer Science, University of Toronto, 40 St. George Street, Toronto, Canada\\
$^{5}$Department of Electrical \& Computer Engineering, University of Toronto, 10 King’s College Road, Toronto, Canada\\
}

\date{Accepted XXX. Received YYY; in original form ZZZ}

\pubyear{2015}

\begin{document}
\label{firstpage}
\pagerange{\pageref{firstpage}--\pageref{lastpage}}
\maketitle

\begin{abstract}

We investigate the focal plane wavefront sensing technique, known as Phase Diversity, at the scientific focal plane of a segmented mirror telescope with an adaptive optics (AO) system. We specifically consider an optical system imaging a point source in the context of (i) an artificial source within the telescope structure and (ii) from AO-corrected images of a bright star. From our simulations, we reliably disentangle segmented telescope phasing errors from non-common path aberrations (NCPA) for both a theoretical source and on-sky, AO-corrected images where we have simulated the Keck/NIRC2 system. This quantification from on-sky images is appealing, as it's sensitive to the cumulative wavefront perturbations of the entire optical train; disentanglement of phasing errors and NCPA is therefore critical, where any potential correction to the primary mirror from an estimate must contain minimal NCPA contributions. Our estimates require a one-minute sequence of short-exposure, AO-corrected images; by exploiting a slight modification to the AO-loop, we find that 75 defocused images produces reliable estimates. We demonstrate a correction from our estimates to the primary and deformable mirror results in a wavefront error reduction of up to 67\% and 65\% for phasing errors and NCPA, respectively. If the segment phasing errors on the Keck primary are on the order of $\sim$130 nm RMS, we show we can improve the H-band Strehl ratio by up to 10\% by using our algorithm. We conclude our technique works well to estimate NCPA alone from on-sky images, suggesting it is a promising method for any AO-system.

\end{abstract}





\section{Introduction}

\import{Sections/}{introduction.tex}

\section{Phase Diversity Methodology}
\import{Sections/}{phaseDiversityMethodology.tex}

\section{Application using an artificial source}
\import{Sections/}{calibSource.tex}

\section{Simulation of NIRC2 on-sky images}
\import{Sections/}{simOnSky.tex}

\section{Application using on-sky images}
\import{Sections/}{NIRC2.tex}

\section{Discussion and Conclusions}
\import{Sections/}{discussion.tex}

\section*{Data Availability}
The data underlying this article will be shared on reasonable request to the corresponding author.
\section*{Acknowledgements}
We would like to thank the reviewer for their helpful feedback on this work; their insight led to several improvements on our technique that we would have otherwise missed. The Dunlap Institute is funded through an endowment established by the David Dunlap family and the University of Toronto. The University of Toronto operates on the traditional land of the Huron-Wendat, the Seneca, and most recently, the Mississaugas of the Credit River; the author is grateful to have the opportunity to work on this land.




\bibliographystyle{mnras}
\bibliography{library.bib} 






% ...


\bsp	
\label{lastpage}
\end{document}

%% file: Sections/introduction.tex
The scientific focal plane exhibits great appeal for wavefront sensing, as it is sensitive to the cumulative wavefront perturbations that directly impact the image quality of the telescope. Traditionally, a significant inhibitor to the performance of an adaptive optics (AO) system have existed in the form of non-common path aberrations (NCPA), whereby wavefront deviations introduced in the non-common path between the wavefront sensor (WFS) and scientific focal plane remain uncorrected by the AO-system. In this context, focal plane wavefront sensing (FPWFS) is critical, providing an offset measurement of the NCPA that can be fed back to the AO-system (see \citealt{Hartung2003} for one of the first implementations, or more recently \citealt{Esposito2020}, employing the use of a Pyramid WFS). Fields such as high-contrast imaging, where extreme (high-order) AO systems are conjoined with coronagraphs, have paved the way for direct imaging of exoplanets; they have also greatly benefited from advances in FPWFS \citep[e.g., ][]{Guyon2004,Sauvage2012,Gerard2018}. In this regime, the technique is usually applied to characterize or control wavefront aberrations at high spatial frequencies and under highly dynamic conditions with respect to the static aberrations traditionally associated with NCPA. 

More recently, the application of FPWFS within AO-systems has been explored to address pupil plane discontinuities, under which we highlight two major regimes. The first is centered around quantifying segmented telescope phasing errors that exist post AO-correction at telescopes such as Keck \citep{Rampy2014,Ragland2016}. The second is with respect to what has been called the `island' or `low wind' effect (LWE, see \citealt{Sauvage2016}), whereby pupil discontinuities un-sensed by the AO-system have been observed during nights with both good seeing and low wind conditions. Contrary to high-contrast applications of FPWFS, these two regimes exist on much slower timescales. Both the island effect and low wind effect demonstrate slow dynamic evolution \citep[on the order of minutes, linked to both seeing and local wind speed, see][]{Milli2018}, while the segmented phasing errors can be thought of as relatively static with respect to these timescales \citep{Ragland2018}. Establishing reliable techniques for the quantification and control of both of these types of wavefront errors is extremely important, and much work has been done in tackling the former, quasi-static errors \citep[e.g., ][]{Sauvage2016,Lamb2017b,N'Diaye2018}. However, the importance of mitigating the segment phasing errors remains and will continue to grow with the upcoming era of giant segmented mirror telescopes (GSMTs), also known as Very Large Optical Telescopes (VLOTs). 

Segmented mirror telescopes such as Keck currently employ robust and reliable phasing techniques, where methods such as broadband phasing \citep[BBP;][]{Chanan2000} have been utilized for more than twenty years. The BBP technique has been the workhorse approach, capable of achieving $\sim30$ nm RMS accuracy and is typically exercised every few months at Keck; such a calibration is done during the day and its relatively low cadence can in part be attributed to the large overheads involved. More recently, there have been advances in employing a narrowband approach, where both accuracy and calibration overheads are significantly reduced \citep{Chanan2018}, however such an approach would still rely on calibrations taken during the day. Despite the phasing accuracy achieved with BBP, on-sky observations demonstrate that Keck exhibits `low order residuals' that clearly exceed $30$ nm RMS accuracy, where broad structure within the AO-corrected PSF of NIRC2 remains persistent and show quasi-static behavior; the sources of these errors are still under investigation, but a significant portion is thought to exist in the form of phasing errors \citep{Rampy2014,Ragland2018}. In light of this, it is clearly desirable to explore methods that can isolate phasing errors within shorter timescales than the current broadband phasing techniques. One potential approach is exploiting the use of an artificial source that sees the primary (given it exists) and using focal plane techniques such as Phase Diversity \citep{Gonsalves82,Paxman92} with high frequency (i.e. on a daily basis). However, in the context of Keck no such source exists; consequently, there has been much work trying to employ Phase Diversity on-sky to quantify Keck's primary phasing errors. 

With the first such on-sky approach conducted by \cite{Lofdahl1998}, whereby the images considered were subject to full atmospheric turbulence, the authors concluded more successful results could be achieved in the presence of AO-corrected images. Indeed, efforts have employed Phase Diversity using the AO-corrected focal plane (and defocused plane) to estimate segmented phasing errors \citep{Ragland2016,vanDam2017,Lamb2017b} as well as quasi-static errors for high-contrast AO-systems \citep{Mugnier2008}. The latter approach utilizes long-exposure images to average out the turbulence and consequently leads to seeing mismatches between focused/defocused images \citep[also see][]{Jolissaint2012}, is somewhat time-consuming, and is currently unable (to the extent of our knowledge) to disentangle primary mirror phasing errors from other quasi-static residuals. Another complication when utilizing AO-corrected images arises in the form of AO-compensated phasing errors, where the AO-system will sense and correct a portion (depending on the amplitude) of primary mirror phasing errors; if left unaddressed, any estimate of the phasing errors will be limited by the degree of AO-correction and their true values will not be reliably quantified. To our knowledge, there is no approach that has addressed mitigation strategies to \textit{simultaneously} overcome these two significant issues: the first involving the difficulties that arise if using two (or more) images within the Phase Diversity algorithm and the second in regards to AO-compensation of phasing errors. 

With these considerations formulated, we now address one of the largest implications and questions we would like to answer. There is clearly a motivation to characterize focal plane wavefront errors from on-sky, AO-corrected images delivered by segmented mirror telescopes; whether the goal be to target NCPA or phasing errors, it is critical that the two be reliably disentangled. For example, if one were to employ on-sky Phase Diversity to quantify segment phasing errors only, the quantification would contain both a contamination from the partial AO-correction of phasing errors mentioned above and from NCPA (the degree of which depending on both the algorithm and the characteristics of the NCPA); a subsequent correction to the mirror segments based on this estimate would yield a phased primary that is subject to this contamination. This phased primary might provide adequate focal plane compensation for the particular AO-system in question, but the newly-phased primary will now contain a signature that will be unwanted for future use of other instruments.

Therefore, the goal of this paper is to outline a technique that reliably and simultaneously quantifies both segment phasing errors and NCPA from focal plane images; the procedure developed is specific to images delivered by an AO-system on a segmented mirror telescope. Given the next generation of GSMTs will heavily employ the use of AO-systems as their primary mode of operations, it is clearly desirable to develop such a tool, especially if these wavefront errors can be provided within a short timeframe. The approach in this paper considers the Keck/NIRC2 AO-system, but can be specifically tailored to be adaptable to any future GSMT with an AO-system. This technique is targeted at obtaining this simultaneous quantification in relatively short timescales (tens of minutes), such that potential correction of its error estimations could potentially be applied faster than any potential evolutionary effects. The approach we adopt entails both a tool development and demonstration of the technique under the context of an end-to-end simulation of a real AO-system, such that a transition to a real telescope can be easily facilitated.

In Section \ref{sec:method}, the general algorithm is described and in Section \ref{sec:calib}, its performance is first considered with an artificial source (i.e. both within the telescope structure and reflected off the primary); this approach would be typically considered for a calibration done during the day. We then assess the algorithm's performance under the effects of an AO-system, where realistic on-sky Keck/NIRC2 images are generated with an end-to-end AO simulator (Section \ref{sec:onSky}). In Section \ref{sec:nirc2}, we subsequently assess the algorithm's performance when: (i) subject to unwanted segment phasing compensation from the AO system and (ii) mitigations of these undesirable effects are addressed. We then discuss the most ideal approach of our method in both the context of Keck/NIRC2 and future GSMTs.

%% file: Sections/phaseDiversityMethodology.tex
\label{sec:method}

The algorithm in our approach is centralized around the Phase Diversity formulation brought forth by \citet{Paxman92}, whereby an objective function is derived defining the maximum likelihood estimate of aberration parameters (i.e. coefficients of a modal basis such as Zernike Polynomials) of an optical system. In particular, we focus on the case where these aberration parameters represent the wavefront residual between the true image and one of a theoretical unaberrated point source. We emphasize this case relies on the object being known and assumed a point source. As such, this only requires one image of sufficient diversity to capture the phase errors in the image; this approach has been employed in the past and is known in the literature as \textit{Phase-Diverse Phase Retrieval} \citep{Ellerbroek1997,Lofdahl1998}. Here, we outline a formulation of our algorithm below in the context of such a scenario.

\subsection{Objective and gradient function}

The general image formation process of a telescope point spread function (PSF) at the detector plane can be mathematically reflected as the square modulus of the inverse Fourier transform of the electric field in the telescope pupil; this scenario is for a point source at a specific wavelength $\lambda$. If the PSF is considered in the context of a detector with $ N \times N$ pixels and $\chi = \{1,2,...,N\} \times \{1,2,...,N\}$, then the PSF, $M(u)$, can be written of the form:

\begin{equation}
    M(u) = \big|\mathcal{F}^{-1}\big\{ A(u)e^{-i\phi(u)} \big\}\big|^2,
\end{equation}

 \noindent where $u \in \chi$, $\mathcal{F}$ is the Fourier Transform, $A(u)$ is the pupil function, and $\phi(u)$ is the phase of the optical system. We can choose to express the phase alternatively in the form

\begin{equation}
    \phi(u) = \sum_{k=1}^{K}\alpha_k \phi_k(u),
\end{equation}

\noindent where $\phi_k(u)$ is typically a set of orthonormal basis functions (e.g. Zernike Polynomials) and $\alpha_k$ are the corresponding coefficients of the basis. In the context of Phase Diversity, a known phase injection $\theta$ is applied (typically in the form of focus), providing an additional known parameter such that the original phase, $\phi$, can be subsequently estimated. The diversity is required to overcome any sign ambiguities in the basis estimation \citep[see][]{Gonsalves82} and must also be of sufficient magnitude to fully estimate the aberrated wavefront (we explore this further in Section \ref{sec:calib}). The equation governing this estimation is derived in \citet{Paxman92} and is also known as the objective function; a slightly modified version of their derivation is

\begin{equation}
\label{eq:paxman14}
    L(f,\alpha) = -\frac{1}{N^2}\sum_{u\in \chi}|D(u) - F(u)H(u)|^2,
\end{equation}
where
\begin{equation}
\label{eq:H}
    H(u) = \big|\mathcal{F}^{-1}\big\{ M(u)A(u)e^{-i\theta(u)} \big\}\big|^2.
\end{equation}

\noindent Here $D(u)$ and $F(u)$ are the inverse Fourier transform of the detected image and object $f$, respectively. $F(u)H(u)$ essentially represents a model counterpart to $D$, including a known injected diversity. When the object is known (i.e. a point source), a single diverse image can suffice for $D$, and the Fourier Transform of $F$ is unity; in this scenario, Equation \ref{eq:paxman14} reduces to a simple minimization between the detected image $D$ and its model $H$. The analytical gradient to this objective function is written below (where we have slightly modified the equation from \cite{Paxman92} to take into account that the object is a point source):

\begin{equation}
\label{eq:paxman22}
    \frac{\delta}{\delta\alpha_k} L_M = -\frac{4}{N^2}\sum_{u\in \chi}\phi_n(u) \text{Im}\big [ H(u) [Z(u)*H(u)^*] \big ],
\end{equation}
where,
\begin{equation}
    Z(u) = \big (H(u)-D(u)\big)^*.
\end{equation}

With this objective function and its analytical gradient now defined, we now aim to utilize an optimization technique to find the solution that best yields a set of aberration parameters, $\alpha_k$, representative of a given single, defocused image.

\subsection{Optimization parameters}

Our algorithm supplies the objective function and gradient to a selection of non-linear optimization techniques, and for this paper we adopt the Broyden–Fletcher–Goldfarb–Shanno (BFGS) quasi-Newton method, which has shown to work well in the past \citep{Lamb2017b,Lamb2018}. An inexact line search is performed to determine the step size of the optimization and the iterations of the algorithm are repeated until a tolerance (for this work, we choose $1e^{-6}$) is achieved. The defocused image supplied to the algorithm (whatever plate scale it may be) is reduced to Nyquist sampling to speed up the computation and the number of pixels defined in the wavefront are 120x120 (any dimension larger than this did not alter our final results by a large margin and so we keep this parameter for the remainder of the paper). As previously stated, the output of our algorithm is an estimate of the coefficients for a modal basis set (of the user's choice) that best respresent the aberrations in the optical system; our algorithm basis can take the form of Zernike polynomials, Karhunen–Lo\`{e}ve (KL) modes, or a customized set of 36 primary mirror segment piston modes defined over the Keck pupil. We reveal (and confirm in later parts of this paper) that using a combination of these last two basis can yield a simultaneous estimate of both phasing errors and NCPA in the context of a point source - either in the form of an artificial source or an on-sky image of a star assumed to be at infinity; for the remainder of this work we adopt such a combination of bases,.

%% file: Sections/calibSource.tex
\label{sec:calib}

We choose to first explore the algorithm in the context of a theoretical artificial source, located within the telescope structure (or beyond) such that it illuminates the primary mirror and appears as a point-source; the implementation of such a source remains outside the scope of this paper and its consideration is only for theoretical purposes. We emphasize a key assumption that this point source is not subject to any atmospheric effects whatsoever. The goal in this Section is to assess the algorithm's ability to simultaneously estimate known input segment phasing errors and NCPA. The motivation of this scenario is twofold: i) to show how beneficial such a source could be in the context of any segmented mirror telescope, and ii) to help refine general algorithm parameters that will then be later used in the context of simulated on-sky images.

\subsection{Generation of images}

\begin{figure}
 \includegraphics[width=0.49\columnwidth]{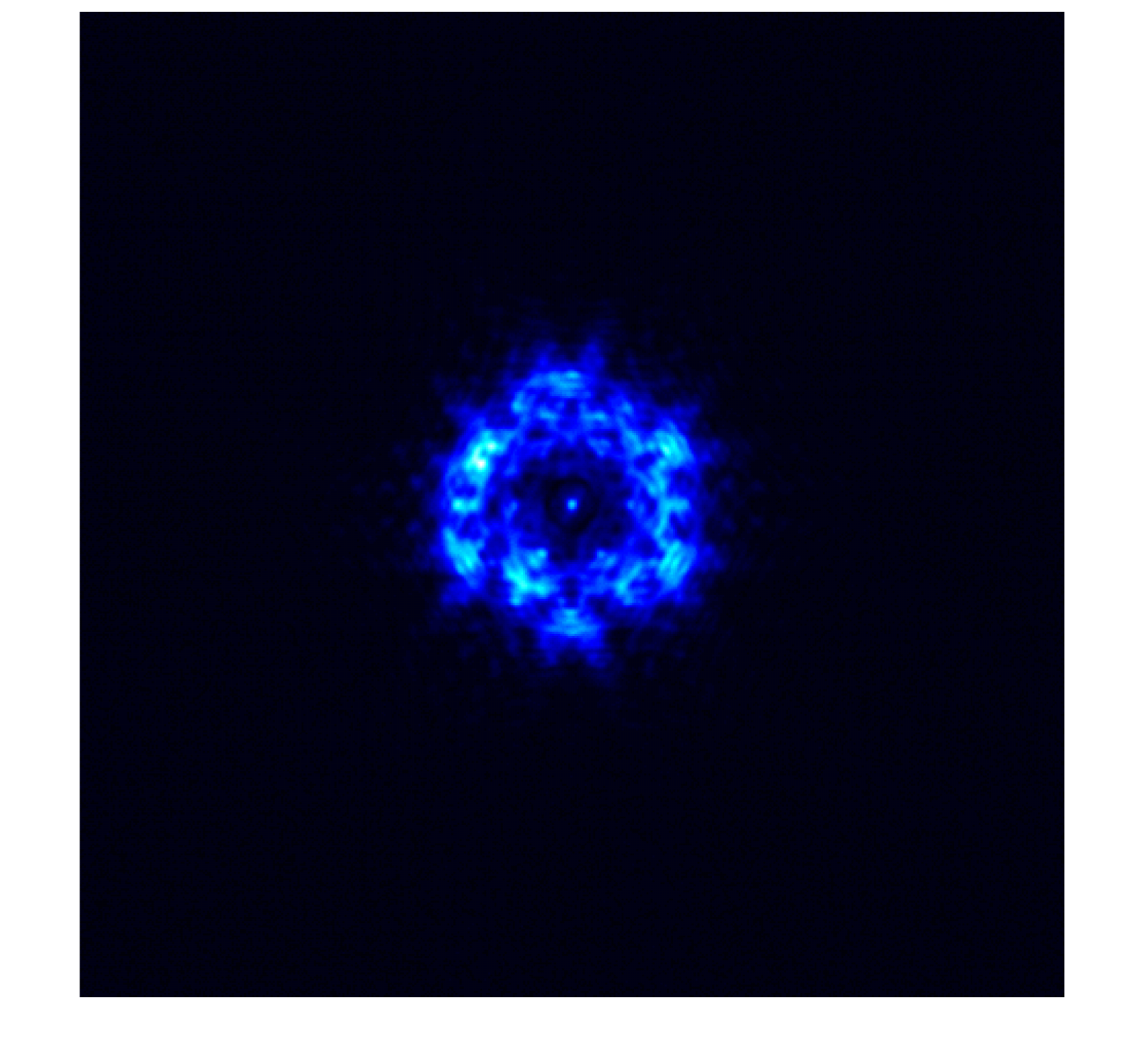} \includegraphics[width=0.49\columnwidth]{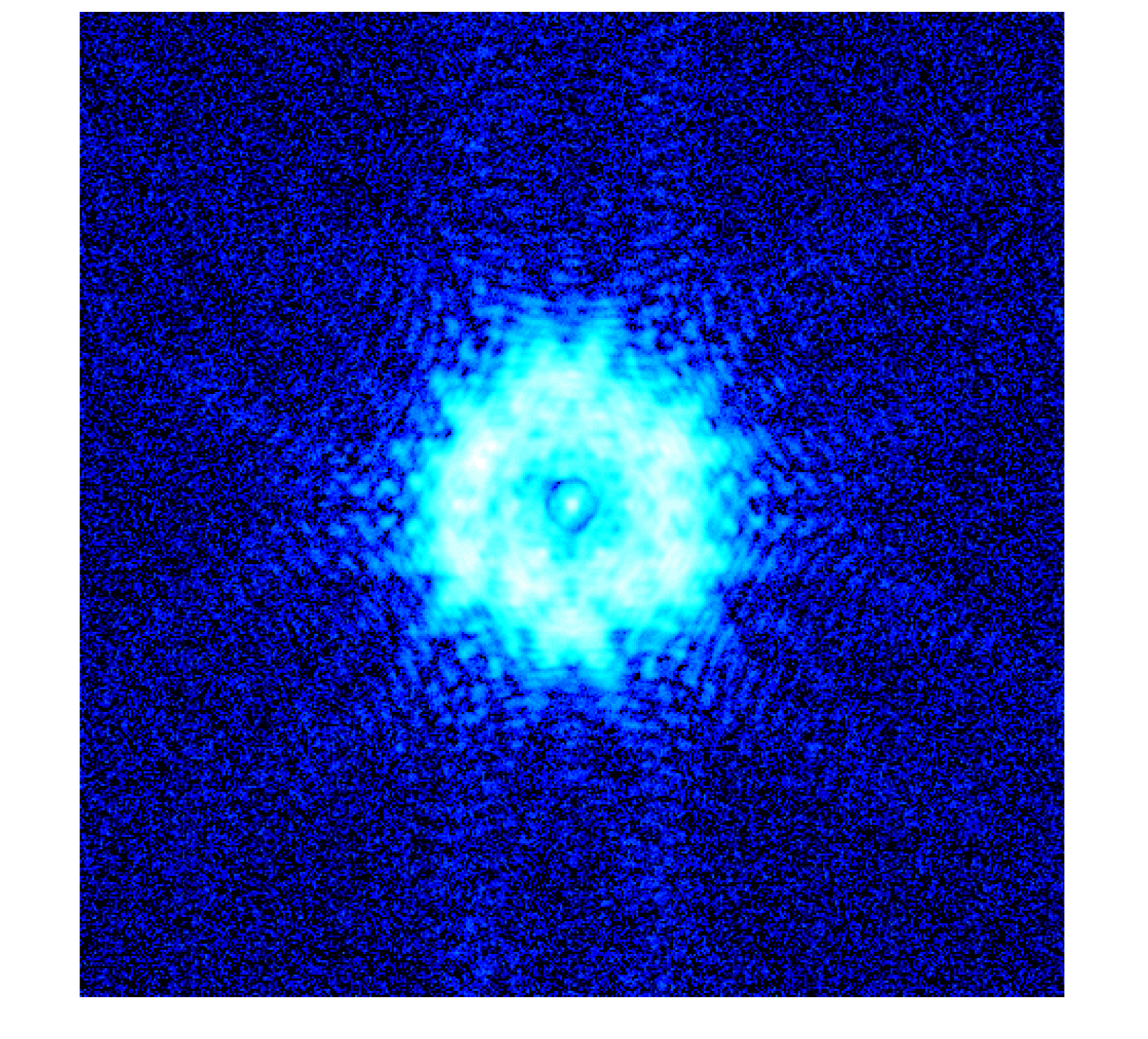}
 \caption{Simulated diverse image (also shown with log stretch on the right) of an artificial point source with $3/2$ $\lambda$ rms defocus, subject to the phasing errors and NCPA shown in Figure \ref{fig:calibPhase} (middle bottom and top). The algorithm uses this image to estimate the coefficients of a model basis representative of both segment piston and KL modes; the results from this estimate are shown in Figure \ref{fig:calibModes}.}
 \label{fig:calibImage}
\end{figure}

In addition to illuminating the primary mirror, the general assumptions about the source is that it is monochromatic (in this work we simulate the Br$\gamma$ narrowband filter offered by Keck/NIRC2) and that the source is not extended (i.e. a true point source). The actual bandpass of this filter is $\sim1.5\%$, due to the relatively narrow filter width of $\sim34$ nm, resulting in a scenario we consider monochromatic. To be consistent with the analysis performed in Section \ref{sec:nirc2}, we establish an artificial detector that mimics the NIRC2 system. Simulated bright, defocused images are then generated subject to $3/2 \lambda$ nm RMS focus (this value is chosen as a starting point, following \citet{Lamb2018} and is then verified in Section \ref{sec:verifyDiversity}). The signal to noise ratio (SNR) on the images is generated such that peak values on the detector are $\sim500$; this regime ensures counts are not extremely high with respect to what one might expect with a real detector (i.e. a HAWAII-2RG) but that a high SNR image is generated to maximize algorithm performance. The read noise was taken as $60$ electrons, which is a largely conservative estimate of a single read from the Keck/NIRC2 ALADDIN3 detector\footnote{\url{https://www2.keck.hawaii.edu/inst/nirc2/ObserversManual.html}}. We select the plate scale of our theoretical detector to roughly match that of NIRC2 in its finest mode (0.01"/pixel), settling on an integer value of 4 pix/FWHM. The general parameters described here are summarized in Table \ref{tab:calib}. Figure \ref{fig:calibImage} shows both a regular and log stretch of the simulated images. We note that no considerations to local turbulence (such as dome seeing) are taken into account and we assume the only aberrations present in the system are due to either primary segment phasing errors or static aberrations due to NCPA.

\begin{table}
	\centering
	\caption{List of parameters used to simulate artificial source images}
	\label{tab:calib}
\begin{tabular}{|l|c|}	
	\hline
	\hline
	Detector read noise & 60 electrons \\
	Image dimensions & 480x480 \\
	Wavelength (narrowband filter) & Br$\gamma$ (2168.6 nm)\\
	PSF sampling (pix/FWHM) & 4 \\
	SNR* & $\sim$500\\
	Exposure time & 10 ms\\
	Diversity focus amplitude & $3/2\lambda$ nm RMS \\
	\hline
	\multicolumn{2}{l}{*Computed at brightest point of defocused image.}\\

	\end{tabular}
\end{table}

\subsection{Simulated segment phasing errors}
\begin{figure}
\hspace{-3mm}
 \includegraphics[width=1.02\columnwidth]{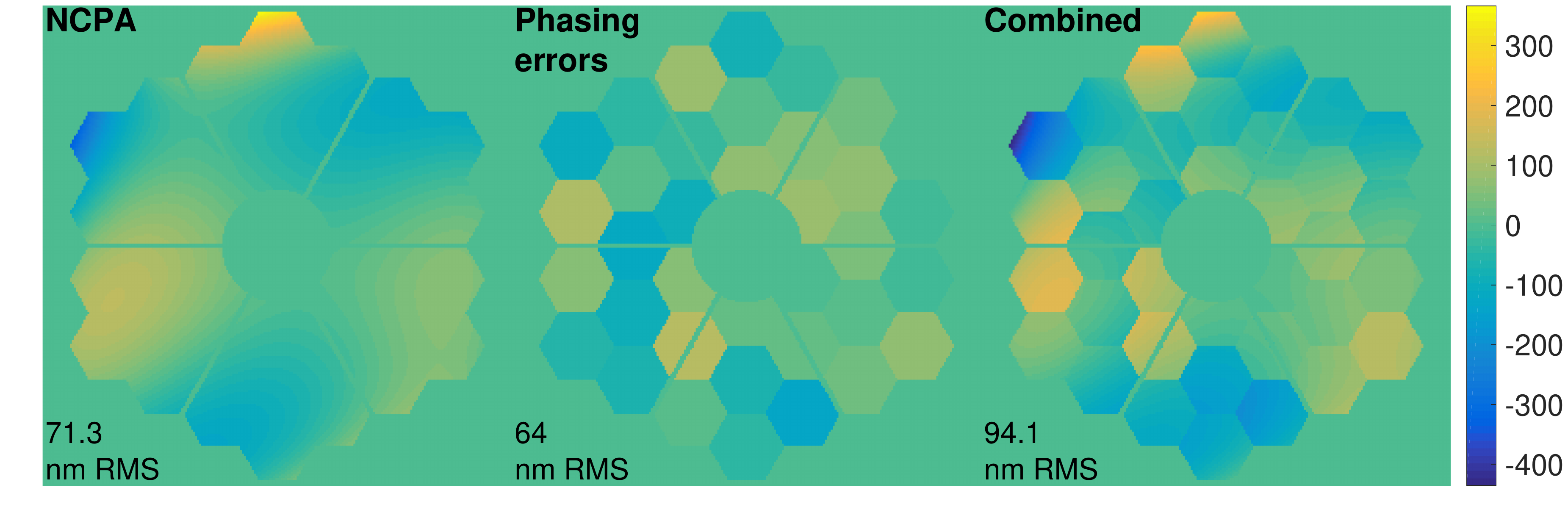}
 \caption{Initial static aberrations considered in our system: NCPA (left), segment phasing errors (middle), and the combination of both (right). }
 \label{fig:initialAberrations}
\end{figure}

The segment phasing errors in this work take the form of differential piston only and do not take into account additional considerations such as \textit{segment stacking} or individual segment tip/tilt. The magnitude of the simulated segment phasing errors are taken such that a random set of 36 segments reflect an RMS wavefront error (WFE) of $\sim64$ nm; this magnitude is considered representative of what is typically observed at Keck, where it is thought at minimum $60$ nm RMS piston errors reside after phasing through edge sensing \citep{Ragland2018}. The wavefront generated from these assumptions is shown in Figure \ref{fig:initialAberrations} (top middle) and is adopted for the remainder of this paper; any estimates of the segment phasing errors performed by our algorithm are always subtracted from this `truth' wavefront and the subsequent RMS computation of this residual serves as our performance metric.

\subsection{Non common path aberrations}
One of the goals of this paper is to show that the algorithm can disentangle segment phasing errors from static aberrations (such as NCPA). As such, relatively large ($\sim73$ nm RMS) NCPA are generated with a $1/\nu^2$ power law \citep[following][]{Dohlen2011} in the form of the first 15 Zernike polynomials; this number of modes was motivated by the fact that estimates of the NIRC2 NCPA \citep{vanDam2017} exhibit mainly low order modes, and 15 modes should suffice for the demonstration purposes of this paper. The magnitude of the NCPA was chosen to be slightly larger than the segment phasing errors, such that a simultaneous estimation of both phasing errors and NCPA will be performed in regimes of similar magnitudes; we note this NCPA magnitude is not entirely unreasonable for what is typically observed on an AO-system. The wavefront representing our selected NCPA is shown in the left of Figure \ref{fig:initialAberrations}. As with our simulated segment phasing errors, this wavefront is adopted for the remainder of the paper and one of the algorithm performance metrics is the residual with respect to this wavefront.

\subsection{Basis selection and algorithm performance}
As described in Section \ref{sec:method}, the algorithm estimates the coefficients to a modal basis; these coefficients can then be subsequently projected onto a wavefront and compared with the known input `truth' wavefronts. The basis estimated for the entirety of this work is a combination of 36 primary mirror segment piston modes and 66 KL modes. The KL basis was chosen due to its ability to generate modes corresponding to shapes explicitly producible by the DM \citep[i.e. by construction, see][]{Esposito2010,Heritier2018}; if the NCPA can be reliably disentangled and estimated, then their compensation could easily be achieved with the DM under this scenario. The number of KL modes was chosen to span a large enough length to overcome any aliasing effects from the estimation of the NCPA wavefront comprised from the 15 Zernikes modes; its specific number (66) pertains to the number of modes contained within 8 radial orders of Zernike modes, which we take to be a reasonable cutoff for our aliasing assumptions.

\begin{figure}
 \includegraphics[width=\columnwidth]{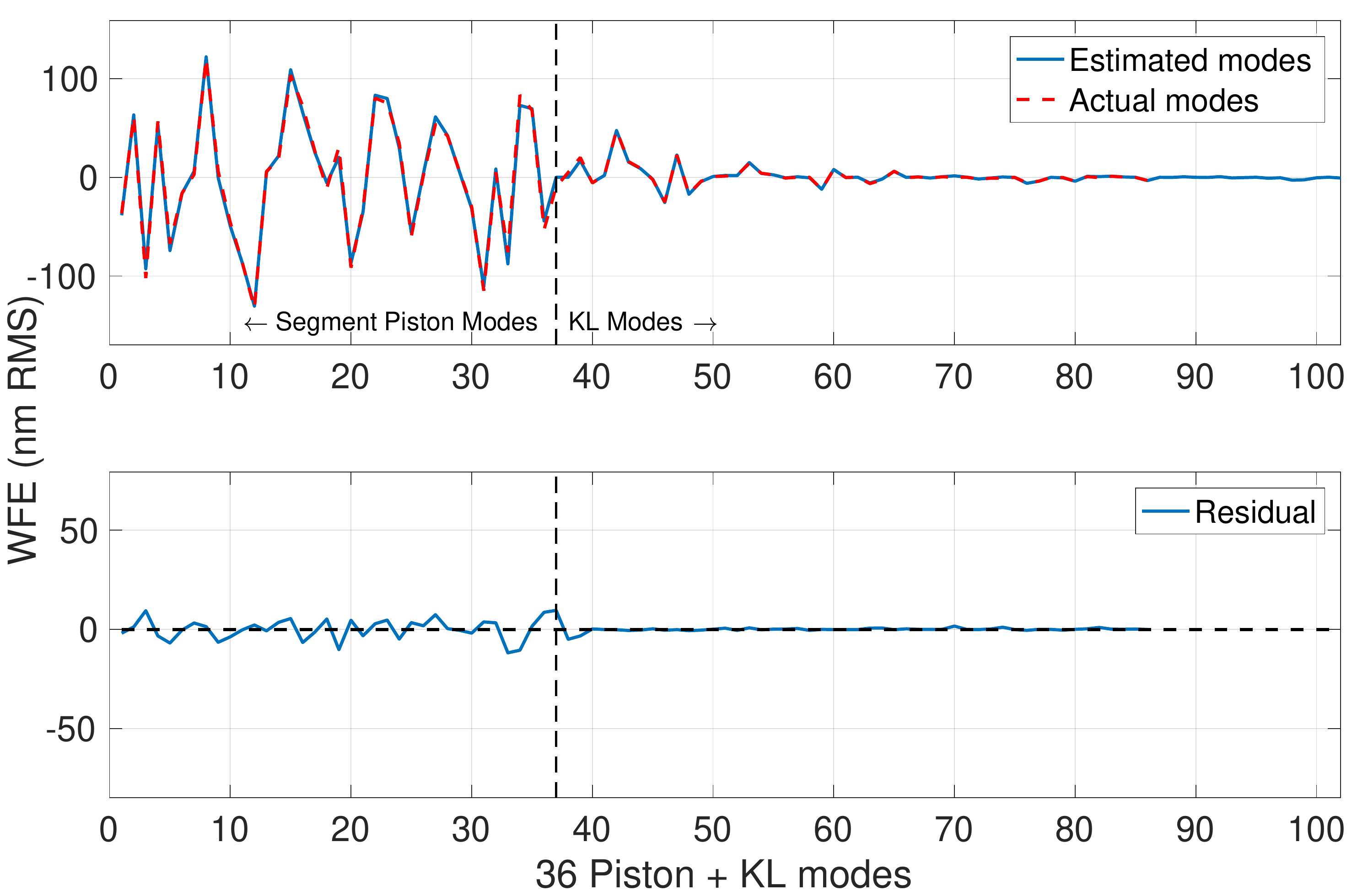}
 \caption{Phase Diversity modal coefficient estimation from our segment piston + KL mode basis in the context of using an artificial source. There are 36 piston modes and 66 KL modes (equivalent to the number of Zernike modes corresponding to 8 radial orders). The estimated modes (blue) agree with the actual modes (red) and their residual is shown in the lower plot. The successful modal recovery and (therefore disentanglement of phasing errors and NCPA) is clearly demonstrated by the lower plot of the residuals; the error in the estimate is shown to be within a few nm RMS WFE in Figure \ref{fig:calibPhase}.}
 \label{fig:calibModes}
\end{figure}
\begin{figure}
\centering
\captionsetup[subfigure]{labelformat=empty}
\hspace{-3mm}
      \begin{subfigure}[b]{0.49\textwidth}
         
         \caption{Phasing errors} \vspace{-0.8mm}
         \includegraphics[width=\textwidth]{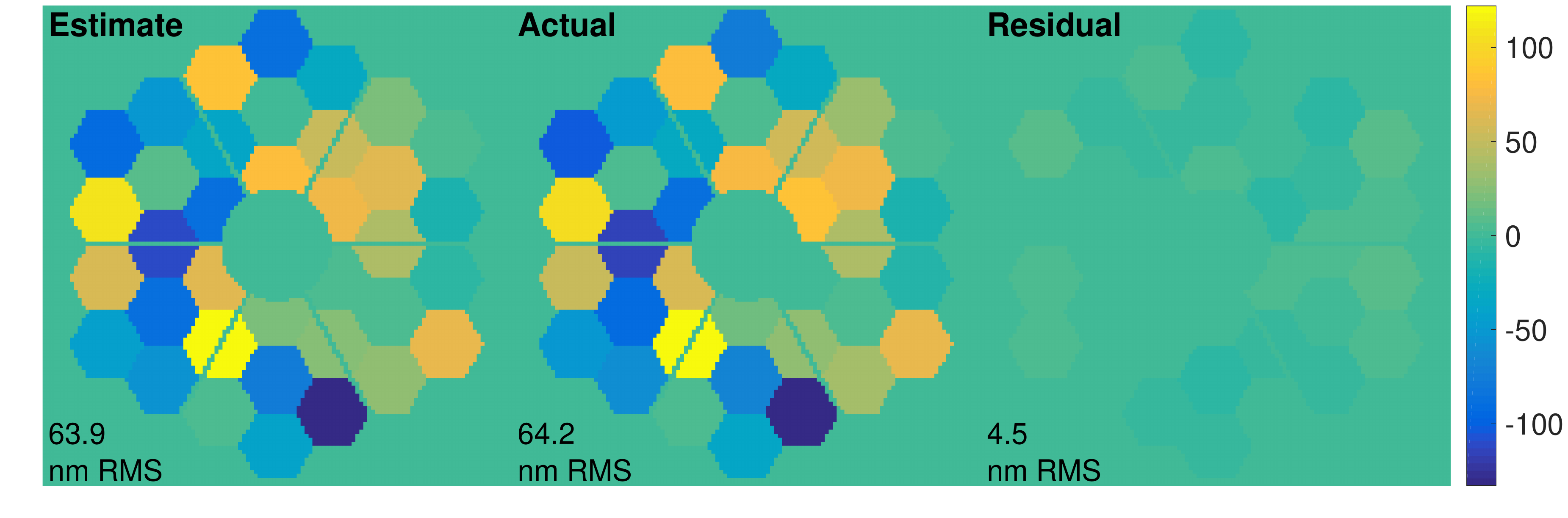}
     \end{subfigure}
     ~
     \vspace{-3mm}
     \\
    \hspace{-3mm}
          \begin{subfigure}[b]{0.49\textwidth}

         \caption{NCPA} \vspace{-0.8mm}
         \includegraphics[width=\textwidth]{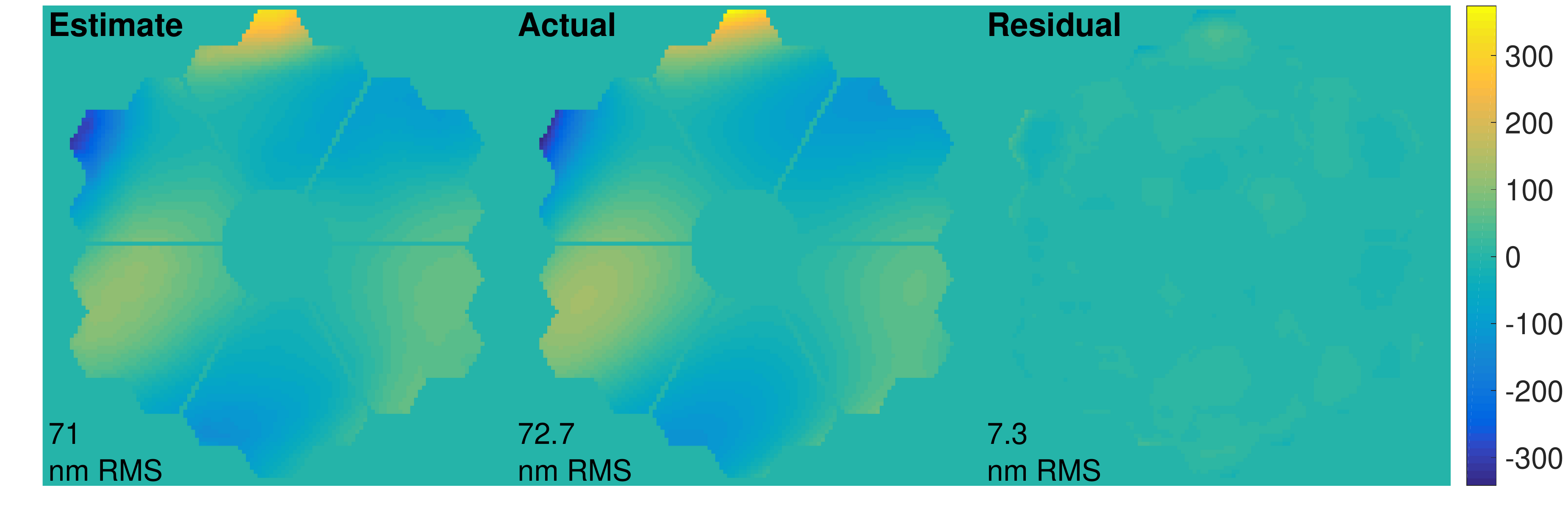}
     \end{subfigure}
     ~

        \caption{Simultaneous estimation of both segment piston errors and NCPA, derived from the projection of estimated modal coefficients as determined from Figure \ref{fig:calibModes}; from left to right for both: estimated phasing errors, input phasing errors and residual between the two (tip/tilt removed).}
        \label{fig:calibPhase}
\end{figure}

The modal estimation of this joint basis from our algorithm is shown in Figure \ref{fig:calibModes}. The 36 element piston basis is clearly disentangled from the input NCPA as indicated by the nearly negligible residual shown in the lower panel of the figure. The modes from this estimation can be projected into wavefront form and subsequently compared with the true input wavefront maps, as demonstrated in Figure \ref{fig:calibPhase}; in this figure, it can be seen that the estimate, truth and residual wavefront is in excellent agreement for both the segment phasing errors and NCPA.

\subsection{Determination of ideal diversity}
\label{sec:verifyDiversity}
\begin{figure}
 \includegraphics[width=\columnwidth]{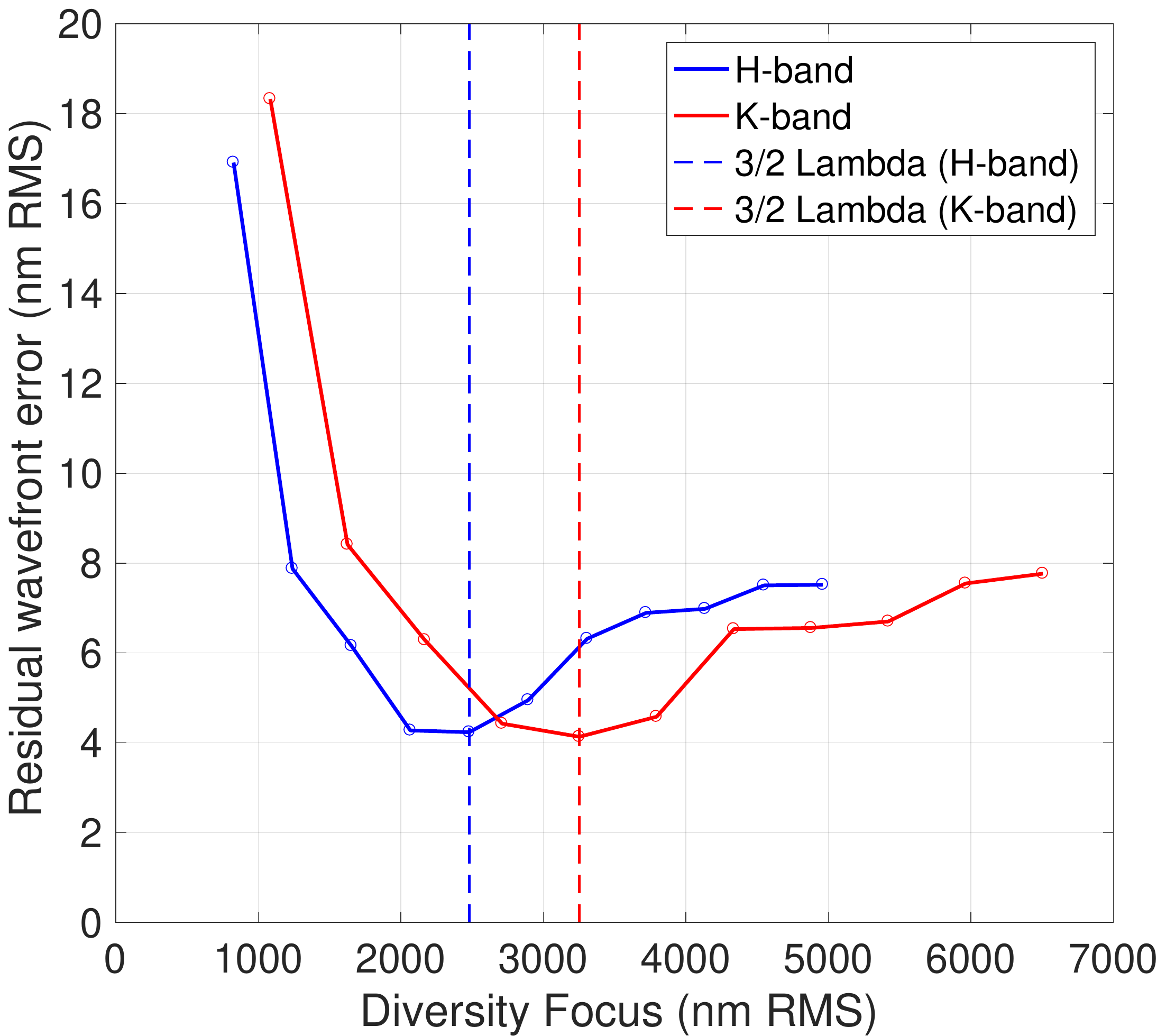}
 \caption{Algorithm performance as a function of focus diversity amplitude. The residual wavefront error (y-axis) was taken as the difference between the estimated and true wavefronts of the phasing errors (see Figure \ref{fig:calibPhase}). The vertical dashed lines represents the optimal value of 3/2 $\lambda$ RMS. This procedure was performed for both H and K band data to confirm our initial choice of diversity (3/2 $\lambda$) for single image Phase Diversity, which was adopted from \citet{Lamb2018}.}
 \label{fig:diversity}
\end{figure}

Using this framework, a range of focus diversity amplitudes was explored to verify our choice of $3/2 \lambda$ nm RMS. The residual wavefront was computed with respect to the input segment phasing errors and for two different wavelengths: narrowband Fe (1.64 micron) and narrowband Br $\gamma$ (2.168 micron). In both scenarios, we confirm our initial choice of focus diversity is ideal and Figure \ref{fig:diversity} demonstrates these results.

The investigation of our technique under the considerations of a reliable artificial source demonstrates that both segment phasing errors and NCPA can be reliably recovered from a single, short exposure focal plane image. While this scenario is not realistic for the current Keck telescope, it is worth exploring given such a source is available on future segmented mirror telescopes. For the remainder of the paper, we focus the application of our technique on a solution that does not rely on an artificial source but rather on AO-corrected, on-sky images; this method can be applied both to the Keck/NIRC2 system and to any future segmented telescope system employing an AO-system.

%% file: Sections/simOnSky.tex
\label{sec:onSky}
We now consider our algorithm in the context of an on-sky scenario of the Keck/NIRC2 system. We emphasize several key considerations in our approach:

\begin{enumerate}
    \item That AO-corrected images of a point source is fundamental to our on-sky approach; outside of this regime the algorithm is not capable of estimating the extremely large wavefront errors of the combined phasing errors and uncorrected atmospheric turbulence. This finding was made by \citet{Lofdahl1998}, and was also verified through initial trials of our algorithm under the same conditions.
    \item A mitigation strategy exists to address seeing evolutionary effects between multiple images used with Phase Diversity. In particular, employing single, defocused images (by assuming the object is a point source) of a star to estimate the phase is a strategy to avoid such a scenario.
    \item The above single, defocused images need to be taken with extremely short exposure (to capture `frozen' residual turbulence; we show how long these exposures need to be in Section \ref{sec:shortImages}) and that they be of sufficient SNR to provide reliable estimation (i.e. AO-correction from very bright natural guide stars).
    \item A moderately large set of defocused images are obtained to provide many phase estimations such that the average estimation will tease out the static errors (phasing and NCPA) from the dynamic AO-residuals.
    \item Finally, that the AO-system itself will partially correct for some of these segment phasing errors, which will in turn present an underlying signature in the Phase Diversity images that ultimately limit the algorithm's ability to estimate the true phasing errors.
\end{enumerate} 

It is clear an approach needs to be developed that addresses all of these considerations, which we outline here. We note this approach is modelled in part after the work developed by \citet{Lofdahl1998}, with key advancements here that pertain to Phase Diversity images considered within the influence of an AO-system.

\subsection{End-to-end AO simulation}

\begin{figure*}
\hspace{-10mm}
 \includegraphics[width=2.2\columnwidth]{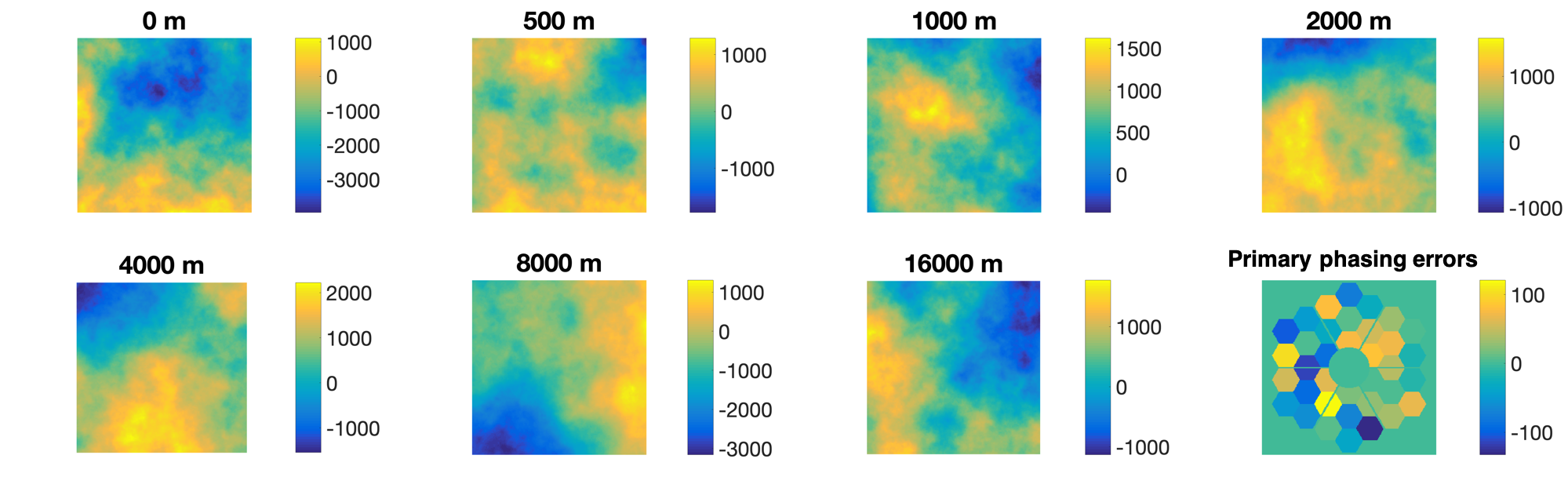}
 \caption{The atmosphere in our AO-simulation is comprised of seven turbulent layers and one static layer; the latter of which is imposed in the form of segment phasing errors and is essentially an additional ground layer (at zero meters, as indicated in the bottom right of the Figure). The colorbar units are in \textit{nm} and the height (in meters) of the layer is shown above each phasemap. The fractional $r_0$ for each layer, along with all other atmospheric parameters are listed in Table \ref{tab:aoParams}. The height and strength of the layers are chosen based on real measurements from Mauna Kea.}
 \label{fig:layers}
\end{figure*}

The AO-corrected, short exposure, defocused images are generated via an end-to-end simulation of the Keck/NIRC2 system using OOMAO \citep{OOMAO}; a detailed list of simulation parameters are summarized in Table \ref{tab:aoParams}, where all AO-related parameters (atmosphere, AO-system parameters, etc.) represent the most up to date values of the true system and site \citep[mostly taken from][with a few alterations from the Keck AO group; priv. communication]{vanDam2004}. The natural guide star (NGS) is simulated on-axis, at a zenith angle of zero; larger angles could be simulated, but for the purposes of this work, our priority is to demonstrate the general feasibility of the algorithm. The NGS magnitude was selected to be of sufficient brightness such that a defocused image pertaining to $3/2\lambda$ nm RMS yielded a relatively high SNR at its brightest point ($>500$); we found a magnitude of $R=4$ was sufficient. We emphasize here that saturation on the detector should not be an issue for such a bright star, as the images will be subject to significant focus, spreading the light over a large portion of the image. We do note, however, that careful considerations would need to be taken with respect to saturating the WFS detector in such a scenario (i.e. use of an ND filter). To incorporate segment phasing errors from the telescope pupil, we include them as an additional static ground layer in our atmosphere model, as shown in Figure \ref{fig:layers}. It is important to note these phasing errors consist of differential piston only, and contain no additional contributions from both mirror segment tip/tilt or stacking (staying consistent with Section \ref{sec:calib}). While this may not be perfectly reflective of the true nature of Keck's phasing errors, it provides a good first order encapsulation of segment phasing errors (assuming they are the dominant contributor) and lay important ground work for future tests that could consider these additional factors. For simplicity, we opt for zonal control \sout{(e.g., instead of KL modal control)}, and filter modes corresponding to a conditioning number of 300; this results in only a few dropped modes corresponding to piston and waffle, due to the relatively small coupling ($\sim 14\%$) of the actuators on the Keck DM. In the following Sections, we will show that KL modal control can be beneficial in some scenarios. The simulation is run with 1 frame of lag error, which is slightly optimistic but should not impact the general methodology of our algorithm. 

\begin{table}
	\centering
	\caption{List of parameters used in the AO simulation.}
	\label{tab:aoParams}
\begin{tabular}{|l|c|}	
    \multicolumn{2}{c}{\textbf{General parameters}}\\
	\hline
	\hline
	$r_0$&16 cm\\
	$L_0$& 50 m\\
	NGS magnitude & R = 4 \\
	Pupil sampling & 240 pix\\
	Telescope diameter & 10.49 m\\
	Telescope FOV & 80"\\
	Num. of atm. layers & 7\\
	Fractional $r_0$ [vec.] & [0.517, 0.119, 0.063, 0.061, 0.105, 0.081, 0.054]\\ 
	Layer heights [vec.] & [0, 0.5, 1, 2, 4, 8, 16] km\\ 
	Wind speeds [vec.] & [6.8, 6.9, 7.1, 7.5, 10.0, 26.9, 18.5] m/s\\
	Wind dir. [vec.] & [0, $\pi/2$, $\pi/4$, $3\pi/2$, $8\pi/3$, $\pi/8$, $\pi$] rad\\
	Seg. phasing errors & 64.15 nm RMS\\
    \hline
	\multicolumn{2}{c}{\textbf{AO parameters}}\\
	\hline
	\hline
	Num. of lenslets & 20\\
	Num. of actuators & 21\\
	Actuator coupling & 0.1458\\
	Actuator pitch & 0.5625 m\\
	Influence function & Gaussian\\
	WFS pix. sampling & quad cell\\
	WFS read noise & 0.4 $e^-$\\
	Sampling time & $1054^{-1}$ s\\
	Min. light ratio & 0.50\\
	Frame lag & 1\\
	Control & zonal /modal (KL)*\\
	Closed loop gain & $0.30$\\
	Conditioning num. & 300\\
	\hline
	\multicolumn{2}{l}{*In some instances KL modal control is optimal.}\\
	\end{tabular}
\end{table}

\begin{figure*}
 \includegraphics[width=1.75\columnwidth]{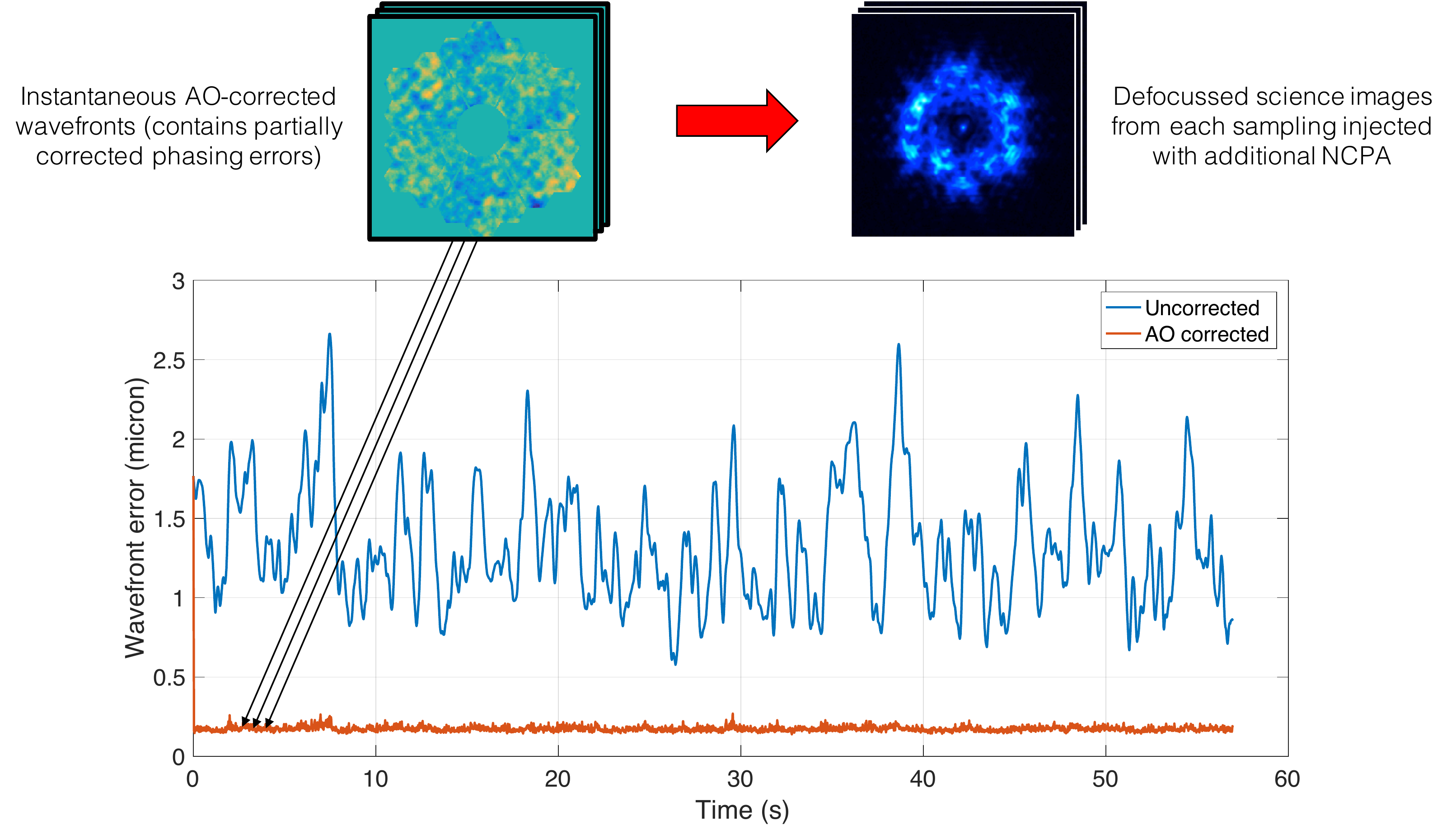}
 \caption{Diagram depicting the on-sky approach used in this paper. The images are acquired in short sequence, each with an exposure time comparable to the coherence time (and thus `frozen'). The approach considers a sequence of images short enough such that the seeing has not significantly evolved but also long enough to acquire a statistically large sample of defocused images. Each of these images is used to estimate both the segment phasing errors and any other quasi-static wavefront (i.e. NCPA) and the average from the sample can be extracted; any dynamic residual wavefront error from the atmosphere should be avoided through this process.}
 \label{fig:method1Loop}
\end{figure*}

\subsection{Generation of on-sky images}
\label{sec:shortImages}
The wavefront residuals from this simulation (which include both the atmosphere and phasing error correction) are combined with the NCPA discussed in Section \ref{sec:calib} to form images simulating the NIRC2 detector. As discussed in Section \ref{sec:calib}, we adopt a read noise of 60 electrons for the detector and also include photon noise. We consider the scenario where the images are under the influence of a `frozen' moment of turbulence, ensuring that there is no `smearing' of the image while the cumulative turbulent projection tracks across the telescope pupil. This scenario is approximated by generating an image from an instantaneous wavefront within the AO-loop at K-band (Br $\gamma$ filter, identical to Section \ref{sec:calib}), where we assume the image can be acquired within the atmospheric coherence time; if we consider Taylor's frozen flow hypothesis, the coherence time of a turbulent layer is $\tau_0 = 0.31 r_0/V_0$, where $r_0$ is Fried's parameter and $V_0$ is average velocity of the turbulence \citep{Roddier1981}. Typical values of $V_0 = 10$ m/s and an $r_0 = 0.15$ m at $\lambda = 500$ nm at Mauna Kea result in $\tau_0 = 4.5$ ms; this is in agreement with measured median values of 4-5 ms \citep{Travouillon2009}. Using this framework, but under the consideration of our synthetic images generated with the Br$\gamma$ filter, we calculate the coherence time to be considerably larger in K-band, resulting in $\tau_0 = 19.5$ ms. We, therefore, operate under the assumption that our imaging system can acquire images within this timescale, which is theoretically achievable with the NIRC2 detector. The simulation is run over a time sequence that we assume unaffected by evolutionary seeing, which we take to be $\sim$ 1 minute. We extract the instantaneous wavefront at $\sim$0.3 s intervals to create our defocused images, yielding a reasonably large sample size of 200. A visual diagram of this process is shown in Figure \ref{fig:method1Loop}. The methodology governing the generation of these images will now be applied in Section \ref{sec:nirc2}, where several on-sky imaging scenarios are explored using the framework of our algorithm.

%% file: Sections/NIRC2.tex
\label{sec:nirc2}

The framework of our algorithm established in Section \ref{sec:calib} can now be applied to on-sky images created from Section \ref{sec:onSky}; we combine the components from the two Sections here and assess the reliability of pure phasing error recovery in the context of two scenarios:

\begin{enumerate}
    \item Where the algorithm is applied to the on-sky images and the known input segment phasing errors (injected prior to the AO-loop) are recovered; this is identical to the diagram outlined in Figure \ref{fig:method1Loop}. The images generated in this scenario are subject to unwanted AO-correction of the phasing errors, limiting the abilities of the algorithm to recover these errors.
    \item The application of the algorithm where the AO-loop is slightly modified such that the effects of the unwanted AO-correction mentioned above is mitigated and therefore providing images which can estimate a more reliable recovery of the segment phasing errors.
\end{enumerate}

\subsection {Direct estimation from on-sky images}
\label{sec:method1}
We first consider scenario (i) above and simultaneously estimate 36 segment piston modes and 66 KL modes from a set of 200 defocused images, adopting the basis from Section \ref{sec:calib}. Following Section \ref{sec:shortImages}, a set of 200 images was generated using our simulation of the Keck AO system. The choice in number of images was selected to provide a reasonably large set of `frozen' images that could be realistically recorded, but also exist within a span of time that is not subject to any significant evolution in seeing (i.e. $<$ 1 minute). The mean estimated coefficients of the piston + KL basis from each image in this set is computed and subsequently projected into two phase maps: the phasing errors and the NCPA. The RMS residual between these two maps and their true wavefront inputs is computed and shown in Table \ref{tab:results} in the second row; it is useful for the reader to note that Figure \ref{fig:calibPhase} demonstrates the same approach used in this computation and can act as a visual analogy.

\begin{figure}
\hspace{-3mm}
 \includegraphics[width=1.02\columnwidth]{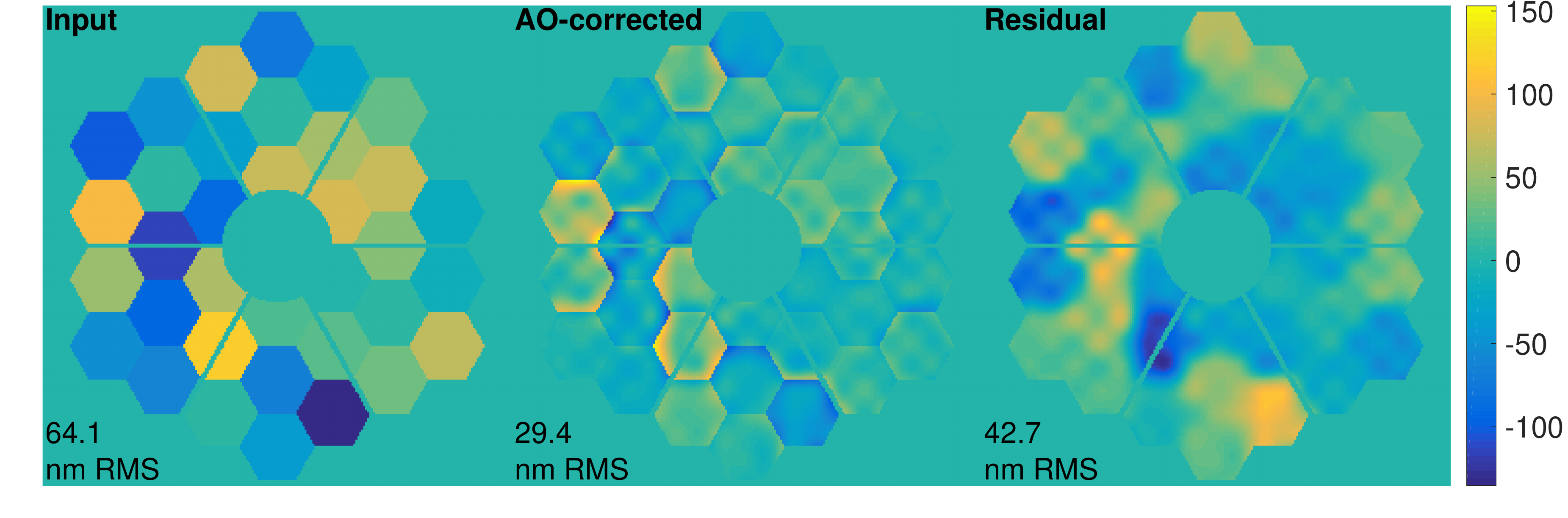}\\
 \caption{Phase maps depicting the effects of NIRC2 subject to segment phasing errors in the form of differential piston only (i.e. phasing errors only and no atmosphere). The input phasing errors (left) show significant features arising from AO-correction (middle), and the residual between the two (right) represents what the AO-system is insensitive to.}
 \label{fig:aliasing1}
\end{figure}

The algorithm is expectedly outperformed by the artificial source scenario, demonstrating the direct effect of being subject to both AO-residual turbulence and unwanted segment phasing correction from the AO-system. The magnitude of the latter effect is clearly visualized in Figure \ref{fig:aliasing1}, where the AO-system loop was closed with the absence of atmospheric turbulence (but keeping the segment phasing errors); the unwanted AO-corrected effects are shown in the right portion of the figure, where the residual between the input phasing errors and AO-correction is shown. This underlying residual will be imprinted upon every defocused image generated by the AO-system, and any subsequent estimation of phasing errors from these images will be hindered by this error. We now address the question: to what degree can we still recover phasing errors and NCPA under this hinderance?

 Although subject to this underlying residual effect, we will show in Section \ref{sec:method3b} that the algorithm is clearly able to disentangle segment phasing errors from NCPA, as was demonstrated for the theoretical source in Section \ref{sec:calib} (see Figure \ref{fig:calibModes} for example); for now, we will operate under the assumption that phasing errors and NCPA can be reliable isolated from one another. The employment of our algorithm results in an estimate of the phasing errors, that when subtracted from the true segment phasing errors, yields a residual of 32.1 nm RMS WFE, demonstrating a 50\% reduction in WFE from the original 64.2 nm RMS. Similarly, the NCPA show a reduction in WFE by 65\%; these results are summarized in the second row of Table \ref{tab:results}.  

To better appreciate the implications of these error estimates, we highlight that in a real scenario one would take the segment piston values of this phasing error estimate and apply it's opposite commands on the actual mirror segments; a perfect application of our estimate would result in a telescope phased to within 32 nm RMS - a 50\% improvement from the original 64 nm RMS. Similarly, because the NCPA can be reliably disentangled, a perfect application of our NCPA estimate would result in a 64\% improvement in WFE reduction.

Given the relatively narrow influence functions of the Keck DM, it is also worth considering whether modal control would be more beneficial. In this scheme, it could allow for a less aggressive form of mode filtering in order to find the optimal set of controlled modes to reduce the unwanted effect of AO-control on the segment phasing errors. Therefore, we perform the same calculation as in the zonal scenario but now with KL modal control. Furthermore, we vary the number of filtered modes for both the zonal and KL modal control schemes to optimize the algorithm performance in \textit{simultaneous} estimation of phasing errors and NCPA. Our optimization metric is the root summed squares of the segment phasing error and NCPA WFEs; the optimal number of filtered modes for each control scheme is the one that minimizes this value, which we find to be 2 and 30 for zonal and KL modal control, respectively. Unsurprisingly, this result re-confirms out choice of filtered modes for the zonal scheme. Filtering 30 KL modes, the algorithm performance now yields segment phasing error residuals of 28.7 nm RMS, $\sim$4 nm lower than its zonal counterpart. However, the NCPA residuals are 26.9, slightly higher than in the zonal scheme. We conclude that KL modal control, with optimized filtered modes, outperforms zonal control when estimating segment phasing errors in this scheme; the results are summarized in Table \ref{tab:results}, third row.

\subsection{Mitigating unwanted AO-correction of segment phasing errors and disentanglement of NCPA}
\label{sec:method3a}

\begin{figure}
\centering
 \includegraphics[width=0.85\columnwidth]{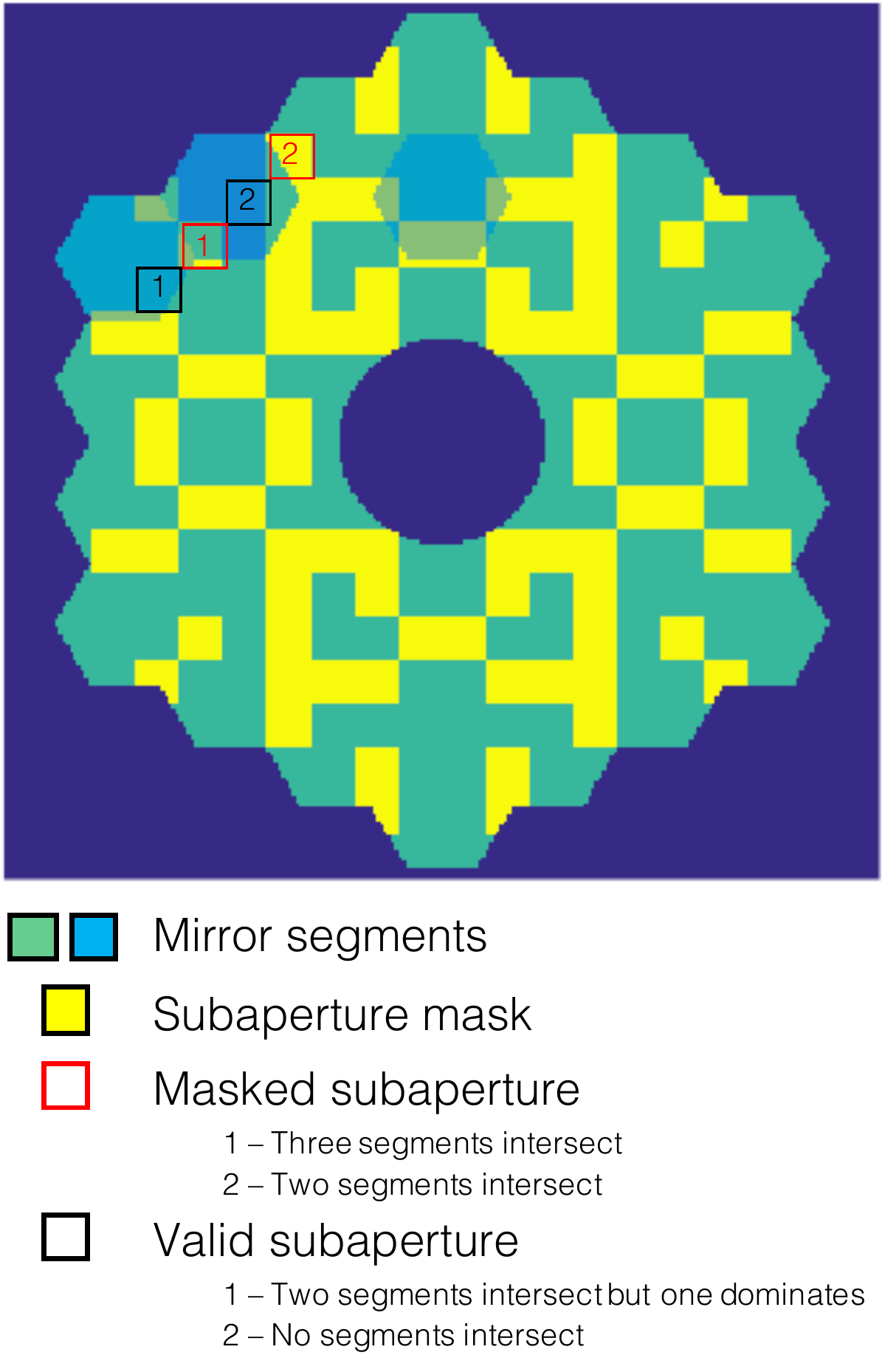}
 \caption{Geometry demonstrating the overlay of our NIRC2 SHWFS mask with respect to the Keck pupil and mirror segments. The unwanted AO-correction from differential piston occurs where multiple mirror segments overlap within a subaperture; the majority of these subapertures (where the effect is considered large) are marked in yellow here and the application of this mask now facilitates a reliable AO-loop closure. In the instances where multiple segments meet but one segment dominates the subaperture (i.e. occupies > 75\%), then we consider the subaperture valid, which in turn improves on the AO system's performance.}
 \label{fig:aliasingMask}
\end{figure}

We now address scenario (ii), where we adopt mitigation strategies to minimize the impact of the AO-correction on the segment phasing errors. The crux of our approach relies on a mask, whereby a particular set of segment phasing sensitive subapertures are intentionally ignored during wavefront sensing. The mask is shown in Figure \ref{fig:aliasingMask} as an overlay of yellow squares on the Keck pupil, with each square representing a suppressed lenslet on the Shack-Hartmann WFS (SHWFS). Also shown in teal are the underlying primary mirror segments. To visualize the intersections between the subaperture mask and mirror segments, we have included additional mirror segments with arbitrary pistons (shown in light and dark blue); black and red outlined boxes indicate a few scenarios where an intersection occurs, but we have chosen to either include or exclude the subaperture in the mask, respectively. The criteria in generating the subaperture mask is if two or more mirror segments intersect within a subaperture on the WFS, then the lenslet is suppressed. A further constraint is that if only two mirror segments intersect, the subaperture is \textit{not} suppressed if one mirror segment dominates in area with respect to the other by a factor of 75\% or more. This value is motivated from our results in Section \ref{sec:method3}, where we rely on AO-loop closure using this mask, and we found this value provided satisfactory loop closure while simultaneously rendering the system relatively insensitive to differential piston (see Figures \ref{fig:aliasing2} and \ref{fig:aliasing3b}).

We now consider two approaches with this mask: one where the mask is directly applied, prior to the generation of the interaction matrix and the loop is subsequently closed; the other approach generates an interaction matrix in the typical way, but then employs an interpolation across this mask in slope space at each iteration within the AO-loop. We discuss the implementation and results of each of these approaches below.

\subsubsection{Masking phasing-sensitive subapertures}
\label{sec:method2}

\begin{figure}
\hspace{-3mm}
 \includegraphics[width=1.02\columnwidth]{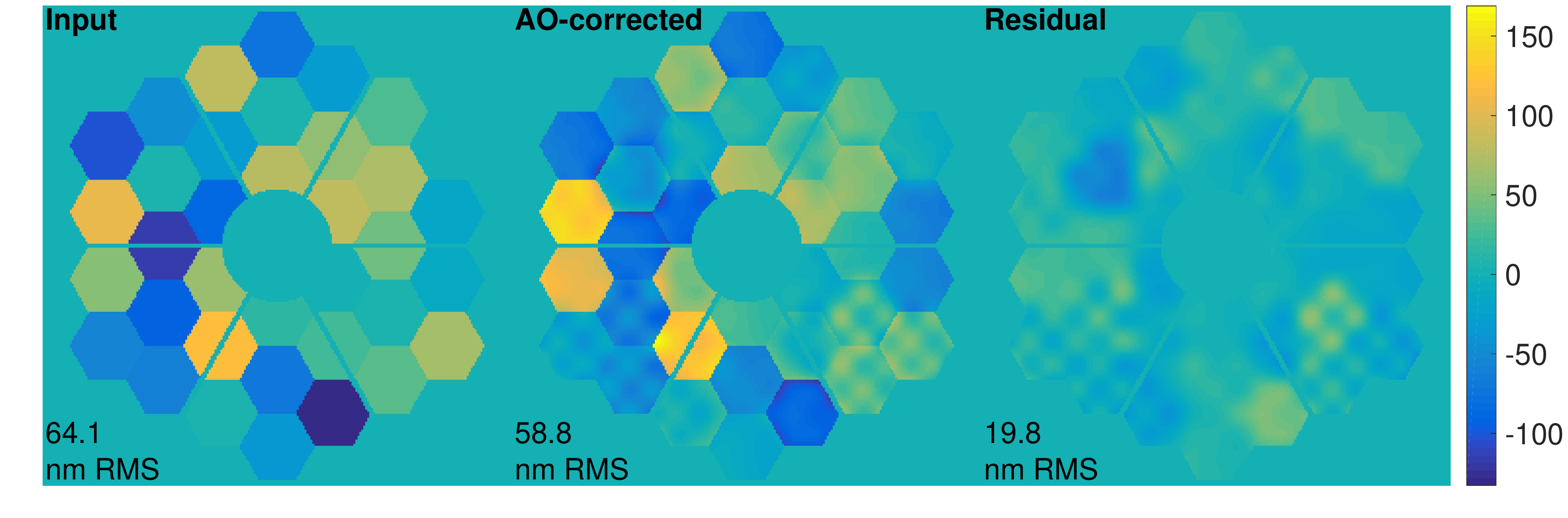}\\
 \caption{Phase maps depicting the effect of AO-correction on segment errors when using the SHWFS mask. Here the interaction matrix is generated with a direct application of the mask and the loop is subsequently closed with no atmosphere. The phase map on the right demonstrates the overall lower residual between the input and AO-corrected wavefront compared to when no mask is used at all (see Figure \ref{fig:aliasing1}), rendering the system to be more insensitive to differential segment piston.}
 \label{fig:aliasing2}
\end{figure}

The direct application of our WFS mask is the most straightforward of the two approaches, where the mask is applied prior to the generation of the interaction matrix. To visualize the effect of the mask, the AO-loop is first closed without the presence of atmospheric turbulence (i.e. phasing errors only) and is shown in Figure \ref{fig:aliasing2}. The right phase map of this figure shows the difference between the input and AO-corrected wavefront, where the system shows a much lower residual error than in the unmasked case (see Figure \ref{fig:aliasing1}) and therefore indicating a lower susceptibility to differential piston. When adding the atmosphere, the best closed loop results were achieved when a larger number of modes were dropped (around 14) compared to the traditional scenario, where only 2 modes were dropped (piston and waffle). We found that while the effect of the mask renders the AO-system to be insensitive to differential piston, it simultaneously degrades the ability of the AO-system to correct for the atmospheric wavefront. To quantify this degradation, the resulting AO-corrected PSF for a short exposure of a bright star yielded a K-band Strehl of about 20 \% (much lower than the un-masked scenario of 79\%). Nonetheless, we wish to still see if our algorithm is capable of estimating the static segment phasing errors and NCPA from a set of 200 images (as described in Section \ref{sec:method1}). Our results indicate the phasing errors can be minimally quantified while the NCPA cannot be reliably estimated. The reported results in the fourth row of Table \ref{tab:results} reflect this finding, where the phasing residual from the truth was 56.9 nm RMS - marginally better than the input of 64.2 nm RMS; the NCPA residual was 87 nm RMS, which is even larger than the initial input of 72.7 nm RMS. Moreover, we found that there was no suitable mask geometry that would yield an AO-correction performant enough to obtain a reliable estimate of the static phasing errors and NCPA.

Despite these initially discouraging results, there are two relatively simple solutions that can help alleviate this problem: (i) considering DM actuators with a larger coupling coefficient and (ii) switching to KL modal control. If the actuator coupling is increased to 0.35, mimicking that of other conventional DMs such as the magnetically actuated mirrors of \textit{ALPAO}, then we find that the performance is greatly improved; this is shown in the sixth row of Table \ref{tab:results}, where the error in both segment phasing errors and NCPA are reduced by 67\% and 56 \%, respectively. We attribute this to the fact that the influence functions with a larger coupling interpolate across the subapertures, and if the coupling is too small (e.g. due to the piezo-stack properties of the Keck DM), then the influence function does not yield reliable interpolation beyond a masked subaperture. The performance in the larger actuator coupling scenario is very promising in the general context of our technique, however it is unrealistic with respect to the Keck/NIRC2 system. 

An alternative approach is considering KL modal control, performing a similar filtering optimization as was discussed in Section \ref{sec:method1}. The idea is that the KL modes can provide a natural interpolation, similar to the effect of using broader influence functions and allow for a better wavefront correction across the masked out subapertures. Indeed, we find from an optimal mode filtration of 100 KL modes that there is a significant improvement from the zonal scenario. Table \ref{tab:results}, row five, summarizes our results in showing this improvement. We note, however, that even with this improvement it does not do as well as if the DM hard broader influence functions; in fact, using this mask under KL modal control does \textit{worse} than if not using any mask at all. Therefore, we now turn to an alternative approach that utilizes the WFS mask that can improve performance and is realistic for the Keck/NIRC2 system.\black

\subsubsection{Interpolation over phasing-sensitive subapertures}
\label{sec:method3b}
The algorithm is now examined in the context of interpolating over our SHWFS mask in 2D slope space. The approach is to first generate a normal interaction matrix (i.e. no application of the mask) and subsequently close the AO-loop; these steps are identical to that of Section \ref{sec:method1}, and the following procedure outlines how the interpolated slopes are achieved. Within the loop, we then apply the mask to the 2D x and y slope maps (thereby masking the unwanted, phasing biased slope measurements) and follow up with a 2D interpolation over each of these maps. We found the best interpolation results from the use of a simple plate metaphor\footnote{John D'Errico (2020). inpaint\_nans \url{(https://www.mathworks.com/matlabcentral/fileexchange/4551-inpaint_nans)}, MATLAB Central File Exchange. Retrieved April 9, 2020.} and the subsequent x and y slopes are extracted and supplied in vector format to the closed loop integrator. We observed that within a few hundred iterations (i.e. $<$ 200 ms) the loop was stable and appeared to provide reliable atmospheric correction. However, after several seconds of loop closure the system diverged. Inspection of the residual wavefront revealed the source of the divergence appeared at the boundaries of the masked subapertures, showing sharp discontinuities in the phase. We postulate this effect arises from the interpolation occurring in AO \textit{residual} slope space, where the relatively large and broad spatial frequencies have been removed by the AO-system (up to the cut-off frequency of the DM). This in turn suggests that valid, unmasked, subapertures will have little to no correlation with respect to their neighbors in terms of large, broad spatial frequencies; given that most of the power in the spectral density profile of the atmosphere is contained within these removed spatial frequencies, there is little spatial frequency information shared between wavefront points at the same resolution of the subapertures. When a subsequent interpolation over the masked subapertures is performed, the method loses its ability to link information across the masked regions and cannot fill in the missing information effectively. This effect is small when done once, but over the course of several thousand iterations of the loop, it diverges in time. 

\begin{figure}
\hspace{-3mm}
 \includegraphics[width=1.02\columnwidth]{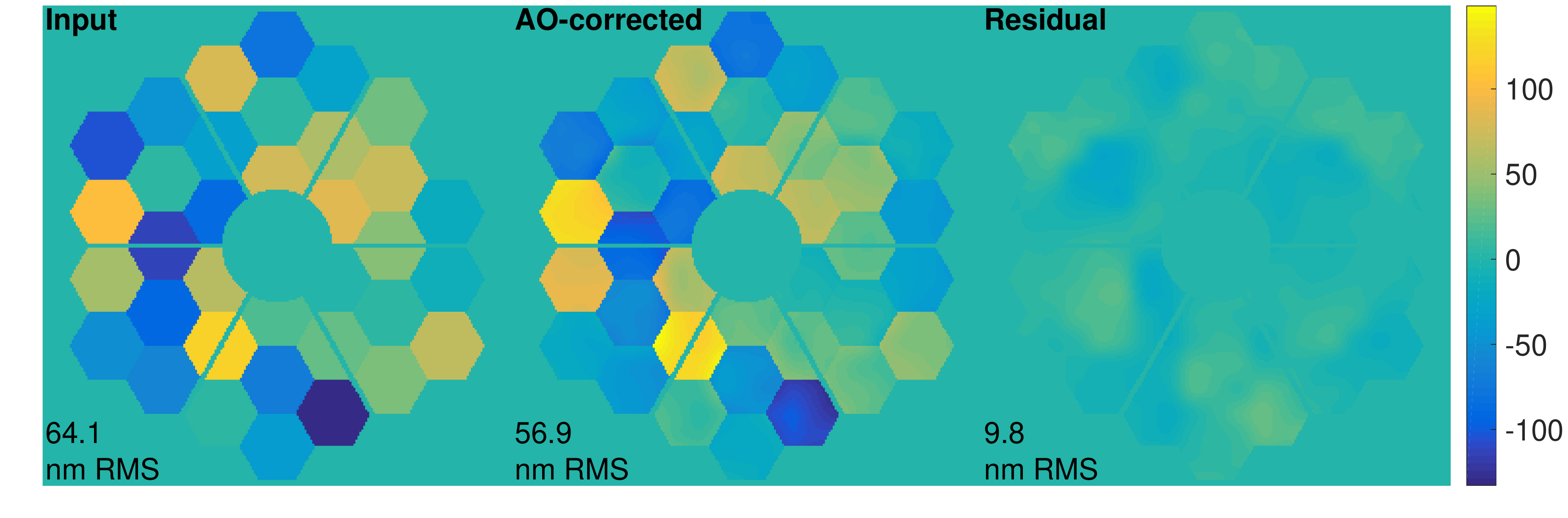}\\
 \caption{The same concept demonstrated in Figure \ref{fig:aliasing2}, but now considering interpolation over the mask in 2D slope space and applying pseudo-open loop control. The residual phase map on the right establishes this approach the best method to suppress the effects unwanted AO-correction of segment phasing errors. This method simultaneously provides reliable AO-correction of the atmosphere such that it can still satisfy the criteria we outlined in the beginning of Section \ref{sec:onSky}.}
 \label{fig:aliasing3b}
\end{figure}

\begin{figure}
 \includegraphics[width=\columnwidth]{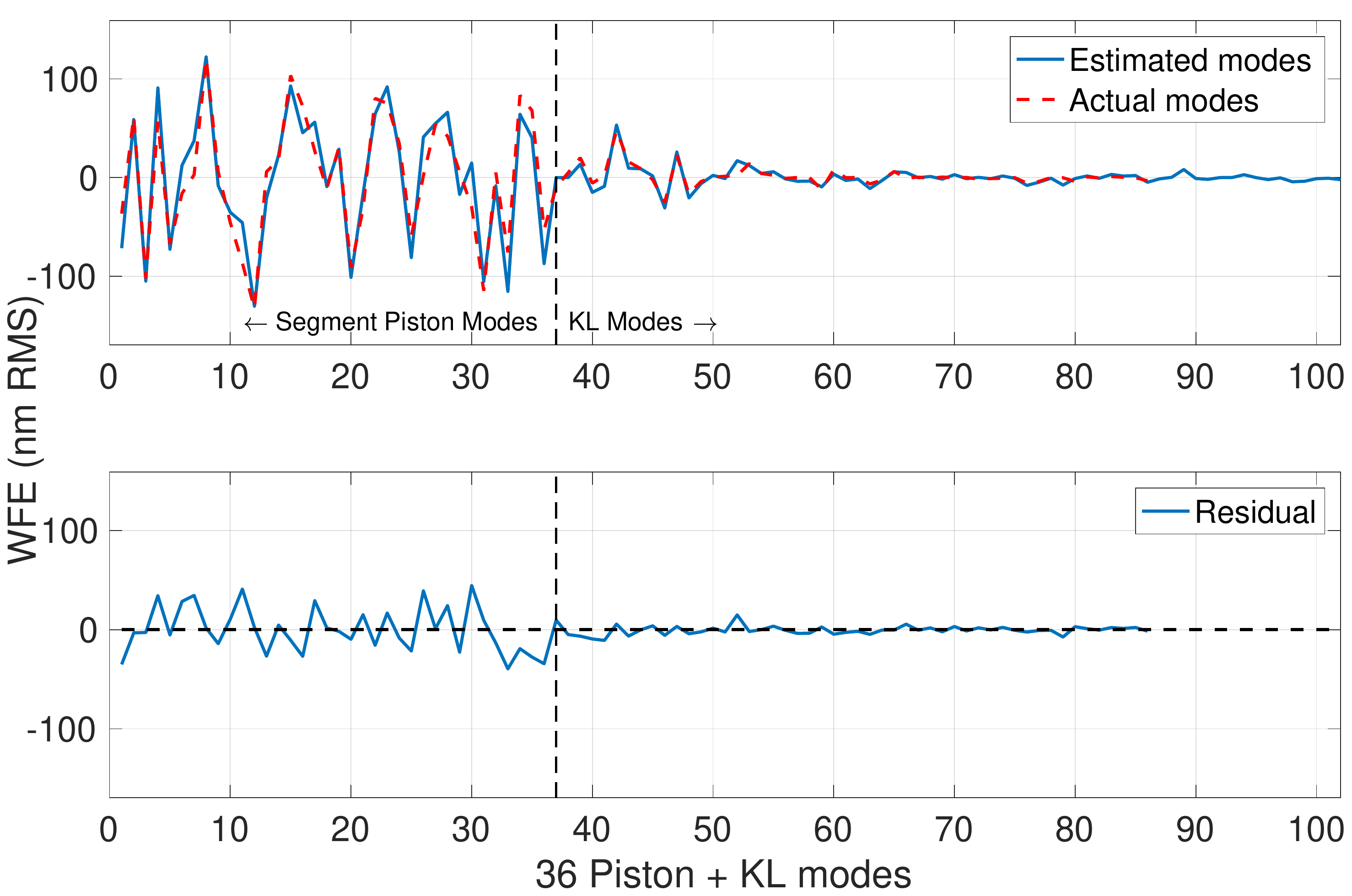}
 \caption{Average modal coefficient estimation from 200 images in the context of POL AO-control employing our WFS mask; the mask allows the AO-system to be rendered relatively insensitive to segment phasing errors while still retaining the ability to reasonably correct the atmospheric wavefront. In the presence of realistic AO-correction, the segment piston modes can clearly be disentangled from the static NCPA component.}
 \label{fig:method3Modes}
\end{figure}

\begin{figure}
\centering
\captionsetup[subfigure]{labelformat=empty}
\hspace{-3mm}
      \begin{subfigure}[b]{0.49\textwidth}
         
         \caption{Phasing errors} \vspace{-0.8mm}
         \includegraphics[width=\textwidth]{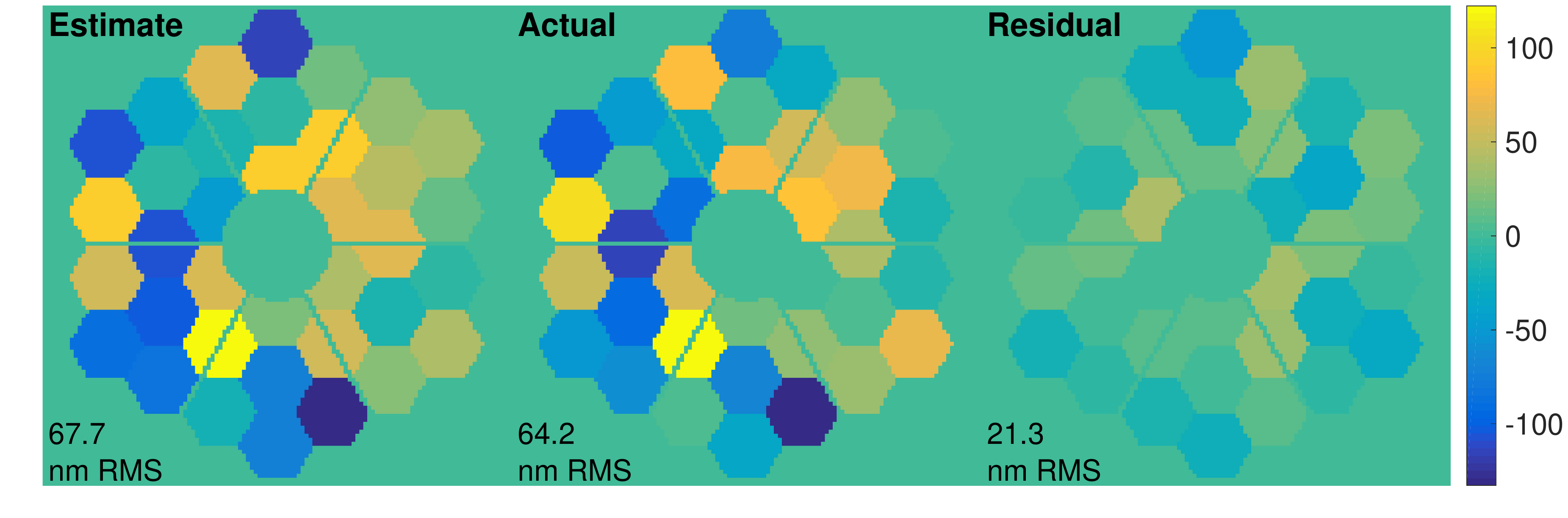}
     \end{subfigure}
     ~
     \vspace{-3mm}
     \\
    \hspace{-3mm}
          \begin{subfigure}[b]{0.49\textwidth}

         \caption{NCPA} \vspace{-0.8mm}
         \includegraphics[width=\textwidth]{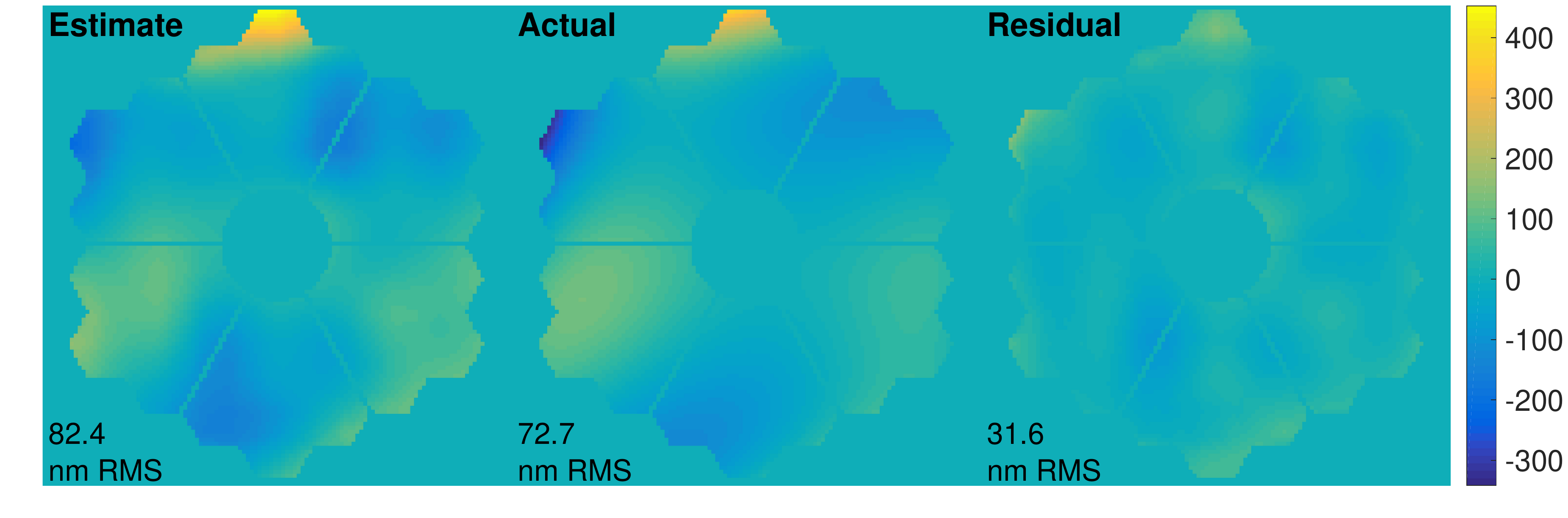}
     \end{subfigure}
     ~

        \caption{Phase projections from the estimated modes in Figure \ref{fig:method3Modes}, showing estimate, truth and residual wavefronts (from left to right) for both segment phasing errors and NCPA (top and bottom). We consider this application the most desirable in the context of the Keck/NIRC2 system.}
        \label{fig:method3Phase}
\end{figure}

A simple approach to validate this theory is to do the exact same analysis in pseudo-open loop \citep[POL,][]{Ellerbroek2003} control, whereby the WFS measurements correspond to the full turbulence, and not the residual, thereby retaining the broad spatial frequencies at scales larger than the resolution of a subaperture on the wavefront. Again, as a visual demonstration, the effect of the phasing error suppression using this approach with the AO-system can be seen in Figure \ref{fig:aliasing3b}; the segment phasing errors of the primary are almost entirely suppressed, down to a level of $\sim$ 9 nm RMS. We now apply the atmosphere and consider this approach, where we find it solves the divergence problem and works remarkably well with our Phase Diversity algorithm. We conduct the same procedure performed in Sections \ref{sec:method1} and \ref{sec:method3a}, and show the modal coefficient estimation of the two basis in Figure \ref{fig:method3Modes} along with their respective wavefront projections in Figure \ref{fig:method3Phase}. It is clear from these two figures that the phasing errors and NCPA are disentangled from each other, even in the context of a closed loop AO-system. Finally, we report the percent error reduction in the last row of Table \ref{tab:results}, where the phasing errors and NCPA were reduced by 66.8 and 56.5\%, respectively. Furthermore, a breakdown of the estimated modes shows the degree to which the two modal basis are disentangled, as demonstrated in the lower portion of Figure \ref{fig:method3Modes}.

It is also worth briefly investigating the impact of KL modal control under this scenario. To gain a sense of how sensitive this method is to modal control, we first consider two scenarios of filtered KL modes in our control scheme: 2 and 30; the former represents the same degree of freedom of controlled modes as the zonal scenario, while the latter was found to be the value that optimizes simultaneous segment phasing error and NCPA estimation, as we found in Section \ref{sec:method1}. Our results indicate a very small increase and decrease in performance for the two schemes of modal control; the residuals for segment phasing errors are 21.8 and 21.1 nm RMS for 2 and 30 filtered KL modes, respectively. Similarly, the residuals for NCPA are 32.6 and 30.7 nm RMS for 2 and 30 filtered KL modes, respectively. These results are very comparable to what was found in the zonal scenario (see Table \ref{tab:results}) and we conclude there is no significant gain to applying KL modal control in this context. We acknowledge that a broader parameter range of filtered modes should be explored in determining the optimal solution, but given our results we anticipate this will have little effect.

 The success of this algorithm has been demonstrated using the combination of a WFS mask and POL control and we emphasize that this approach could be applied to the Keck/NIRC2 system (or any future segmented mirror telescope with an AO-system). This requires the real time controller (RTC) be modified to perform this masking and interpolation at each step in addition to employing POL control. However, we also note that the full implementation of POL need not be applied, where a simple extraction of the full turbulence WFS slopes could be performed via the DM influence functions; this would remove the need to implement POL control and still achieve the same underlying interpolative effect.
 
 \subsection{Performance as a function of on-sky images}
 \label{sec:nImages}
\begin{figure}
 \includegraphics[width=\columnwidth]{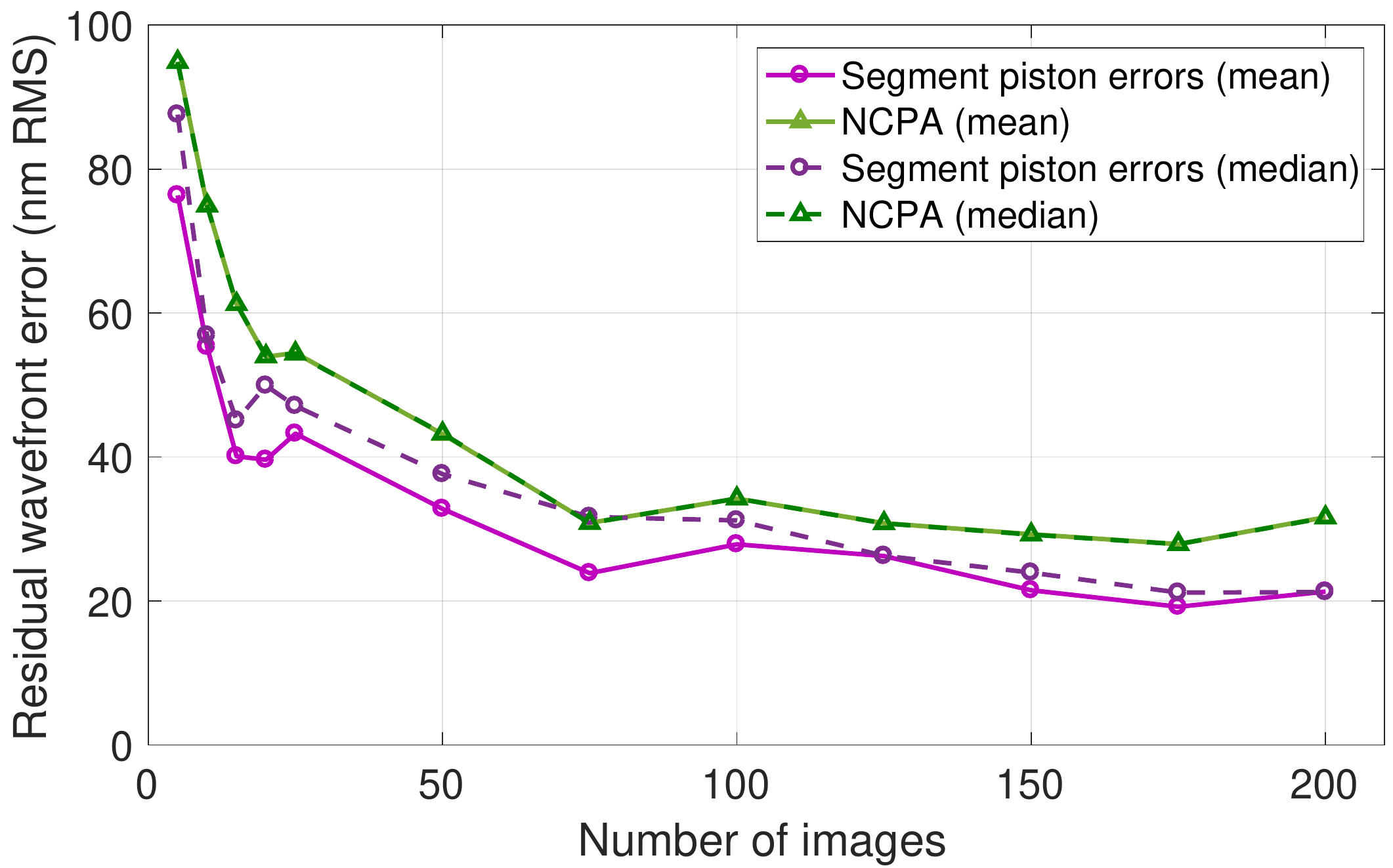}\\
 \caption{Estimation performance of segment piston errors and NCPA as a function of number of average or median combined images. The method considered here is the POL case with an interpolated WFS mask, which we consider the best solution for estimating phasing errors in the context of the Keck/NIRC2 system. The data points corresponding to 200 averaged images is also presented in the last row of Table \ref{tab:results}.}
 \label{fig:nImagePlot}
\end{figure}

 We have established the algorithm as a viable approach to quantify simultaneous phasing errors and NCPA from a set of 200 on-sky, AO-corrected images and we now aim to identify optimal parameters regarding this image set. Specifically, what is the minimum number of images that can yield a reliable estimate and will the median of this set exemplify a better metric than the mean? In effort to answer these questions we repeat the exercise of Section \ref{sec:method3b} (what we consider the best approach for Keck/NIRC2) for a variety of images sets, where we calculate the phase from both the mean and median of the coefficients estimated in these sets. It is important to note that the images generated in this set were still over the entire course of the $\sim$minute of AO-corrected data, and therefore an image set $<$200 images will have images sampled further apart in time. The results from this exercise are easily visualized in Figure \ref{fig:nImagePlot}, where the residual wavefront error for both phasing errors and NCPA (identical to what was calculated in Table \ref{tab:results}) are plotted as a function of number of images used. The wavefront error is computed considering both the mean and the median of the estimated modes. We find a clear trend in wavefront error reduction as the number of images is increased, and our conjecture is that a minimum of 75 images yields consistently desired estimates. Furthermore, adopting the mean over the median shows a clear advantage in segment phasing error estimation when using fewer images, however as the sample size increases, it appears relatively similar to using the median. Interestingly, there is no clear difference when considering the mean or the median when estimating the NCPA. 

\begin{table}
	\centering
	\caption{Estimation performance from 200 averaged on-sky images}
	\label{tab:results}
\begin{tabular}{|l|c|c|c|c|}	
    & \multicolumn{2}{c}{\textbf{Phasing error}} & \multicolumn{2}{c}{\textbf{NCPA}} \\ 
    \multicolumn{1}{c}{\multirow{2}{*}{\textbf{Method}}} & \multicolumn{2}{c}{\textbf{residuals}} & \multicolumn{2}{c}{\textbf{residuals}} \\ \cmidrule(lr){2-5}
    & nm RMS* & $\%^{\dagger}$ & nm RMS & \% \\
	\hline
	\hline
	Artificial source$^{\ddagger}$ & 4.4 & 93.1 & 7.2 & 90.1 \\
	AO correction only (zonal) & 32.1 & 50.0 & 25.7 & 64.6 \\
	AO correction only (modal) & 28.7 & 55.3 & 26.9 & 63.0\\
	AO + WFS mask (zonal) & 56.9 & 11.4 & 87.0 & -19.7\\
	AO + WFS mask (modal) & 37.7 & 41.2 & 33.4 & 54.0\\
	AO + mask (larger IF) & 21.4 & 66.7& 31.9 & 56.1\\
	AO + mask + interpolation & \multicolumn{4}{c}{\hspace{3mm}Loop divergence}\\
	POL AO + mask + interp. & 21.3 & 66.8 & 31.6 & 56.5\\
	\hline
	\multicolumn{5}{l}{*Residual from truth, tip tilt removed.}\\
	\multicolumn{5}{l}{$^{\dagger}$Error reduction \% from the original input.}\\
	\multicolumn{5}{l}{$^{\ddagger}$Using only one image.}\\
	\end{tabular}
\end{table}

\subsection{Impact on general performance metrics}

The ultimate deliverables for an Astronomer employing an AO-system are typically not in the form of residual WFE, but rather Strehl ratio or contrast. We wish to briefly elaborate on the algorithm's performance in terms of these deliverables, and in particular focus on both the H-band Strehl ratio and their corresponding power spectral density (PSD) characteristics (which can be linked to high-contrast for coronographic systems).

We consider two scenarios: one with the phasing errors pertaining to the magnitude discussed throughout this paper (64 nm RMS), and one with double this amplitude, to assess how such an algorithm would perform under these circumstances; we refer to these two scenarios as \textit{conventional} and \textit{substantial} phasing errors, respectively. While the true amplitude of these phasing errors is thought to generally be on the order of $\sim$ 60 nm RMS, there are PSF residuals clearly indicating that the nature of these errors may be much larger in amplitude; for example, \cite{Ragland2018} finds that 120 nm of low order piston modes replicate features seen in the on-sky PSF.

\subsubsection{Conventional and substantial phasing errors}

We outline here a procedure to quantify our performance metrics. The AO-loop is first run with the either the conventional or substantial phasing error introduced to the primary mirror and one minute of short exposure, diverse images are acquired. The algorithm is subsequently run on 75 images (this number was chosen based on the findings from Section \ref{sec:nImages}) to estimate the phasing errors and NCPA, using our favored method of POL control employing an interpolated WFS mask. We reliably disentangle the phasing errors from the NCPA, and we subsequently apply a correction to the primary mirror from our phasing error estimate (assuming a perfect correction can be applied). This disentanglement ensures that the correction applied to the primary will contain little to no NCPA contamination; however, the underlying NCPA will still persist in our final delivered images, as we have made no assumption about their correction. 

The AO-loop is then closed for one second and H-band images are recorded in addition to phase map telemetry. The images are recorded with zero photon or read noise so that a true performance assessment can be obtained; this lack of noise is also why one second of AO-loop data should suffice in an assessment of the Strehl ratio of the PSF. Again, we assume a detector sampling with 4 pixels/FHWM and we sub-pixel shift to yield images with the peak of the PSF lying on a single pixel (employing the sub-pixel centering method described by \citealt{Poyneer2003}). The H-band Strehl ratio is then computed by three methods (all yielding similar results) and our final reported values employ the first of the three: 

\begin{enumerate}
    \item \textbf{Energy normalization:} The on-sky PSF is normalized to the perfect PSF by multiplying it by the ratio between the total sum of the model image and total sum of the actual image, where we have also ensured the two images are centered about the same pixel.  Subsequently, the Strehl ratio can be computed from the ratio of the maximum of this normalized image with the maximum of the perfect PSF (which is unity).
    \item \textbf{OTFs:} The ratio of the summed image OTF with that of the perfect image, where the OTFs are computed over the same box described above.
    \item \textbf{Mar\'echal:} Via the Mar\'echal approximation \citep{Mahajan1983}, which we consider to be valid for the wavefront errors of our application. Due to our simulation, we have access to the telemetry of the residual WFE at each iteration of the loop and the subsequent Strehl can be computed at each iteration via the Mar\'echal approximation; the final Strehl is calculated from the average over all the iterations in the loop.
\end{enumerate} 

\begin{figure}
 \includegraphics[width=\columnwidth]{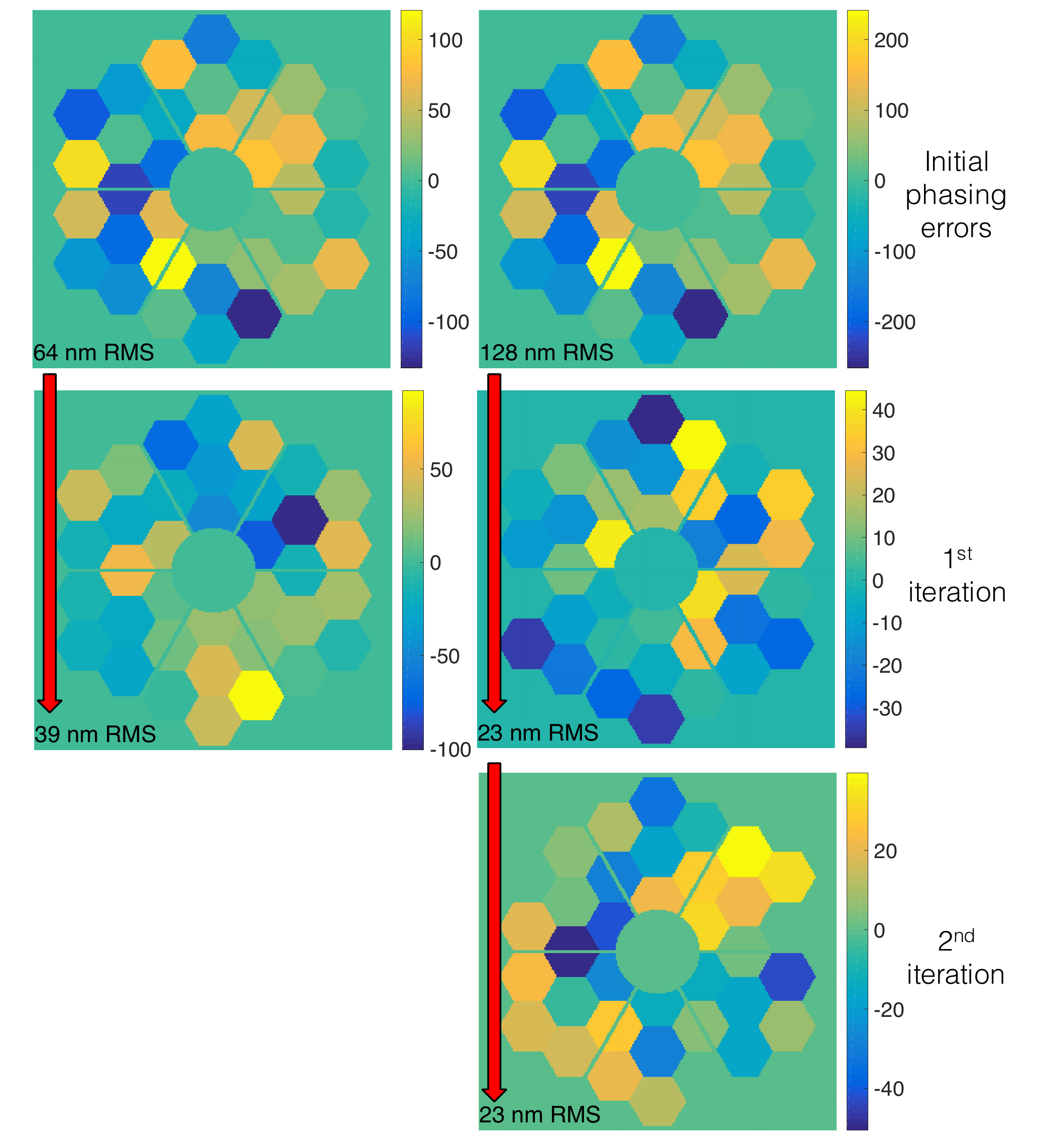}\\
 \caption{Telescope phasing errors used within the simulated AO-loop to compute H-band Strehl ratios. The left column pertains to the use of our algorithm in the context of \textit{conventional} phasing errors, representative of what is thought to be the magnitude of the current piston phasing errors. The WFE reduction if the estimate from our algorithm was applied perfectly (indicated by the red arrow) is also shown; the AO system is run and a one second image is acquired with its corresponding Strehl computed for an H-band imager. The right column depicts a similar scenario, however under the influence of \textit{substantial} WFE, where the initial phasing error amplitude has now been doubled. Our algorithm requires an additional iteration (described in the text) and ultimately achieves the same performance in terms of residual WFE.}
 \label{fig:phasingResiduals1and2}
\end{figure}

The residual WFE reported in Figure \ref{fig:nImagePlot} pertaining to 75 images is analogous to a perfect correction from our PD phasing error estimate; we apply this error and close the AO-loop and report on the H-band Strehl following the prescription above. The resulting Strehl is 62\%, with an initial Strehl of of 61\% (closing the AO-loop on the initial phasing errors alone). While the improvement from this correction appears relatively minimal, we calculated the Strehl of a perfectly phased telescope to be 62\%, demonstrating the correction from our algorithm is close to the near-perfect system. However, this is within the context that the initial phasing errors (of piston only, corresponding of 64 nm RMS) are realistic; the impact of such errors here show a Strehl reduction of only 1\%, and in reality this affect has been observed to be larger. Figure \ref{fig:phasingResiduals1and2} demonstrates the phasing error reduction, as indicated by the red arrows; the phasing errors here were reduced from 64 to 22 nm RMS WFE. We now consider a scenario where we have doubled the input phasing error WFE to 128 nm RMS (i.e. \textit{substantial} phasing errors).

Following the same procedure, but now applying the larger \textit{substantial} primary phasing errors, we find a ground truth residual WFE of $\sim$39 nm RMS - substantially larger than the $\sim$23 nm of the \textit{conventional} phasing scenario. However, if the phase diversity procedure is repeated with a second iteration, feeding the 39 nm RMS WFE as input initial phasing errors, the resulting ground truth residual is further reduced to $\sim$23 nm RMS (see Figure \ref{fig:phasingResiduals1and2}).

\begin{figure*}
\hspace{-3mm}
 \includegraphics[width=2.1\columnwidth]{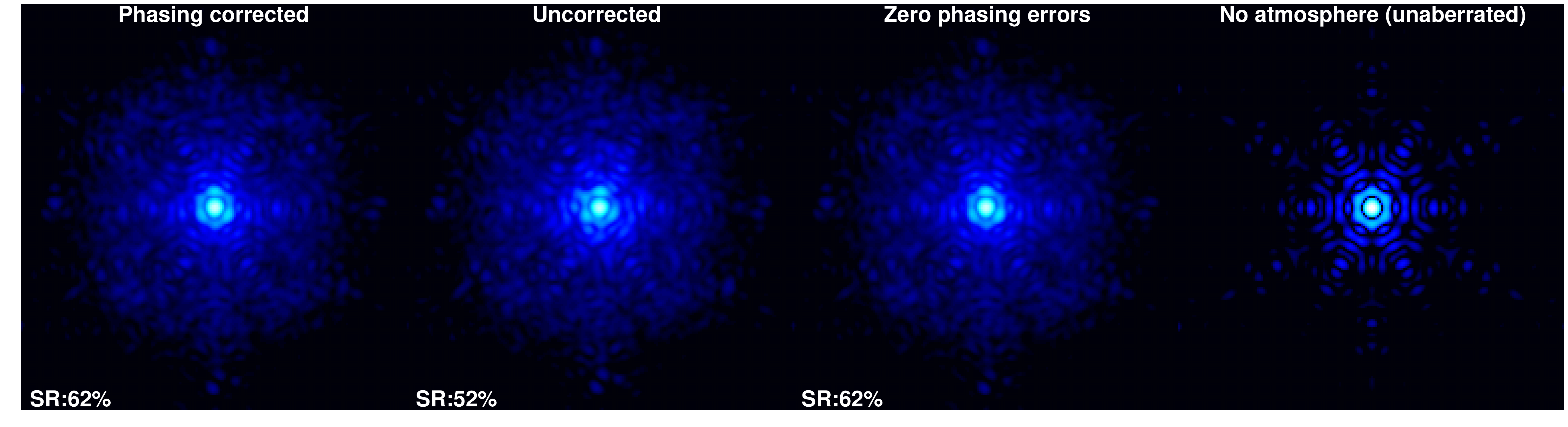}\\
 \caption{Simulated on-sky, 1 second exposure, H-band images of a bright (magnitude 0) star corresponding to substantial initial telescope phasing errors (128 nm RMS WFE) and computed Strehl ratios. These images are taken in the presence of zero photon or read noise to better understand the relative improvement after a phasing error correction. From left to right: resulting image from phasing error correction after two iterations of our algorithm, the uncorrected PSF subject to these phasing errors, the PSF if there were no phasing errors at all and a perfect PSF for reference.}
 \label{fig:strehls}
\end{figure*}

\begin{figure}
 \includegraphics[width=0.5\columnwidth]{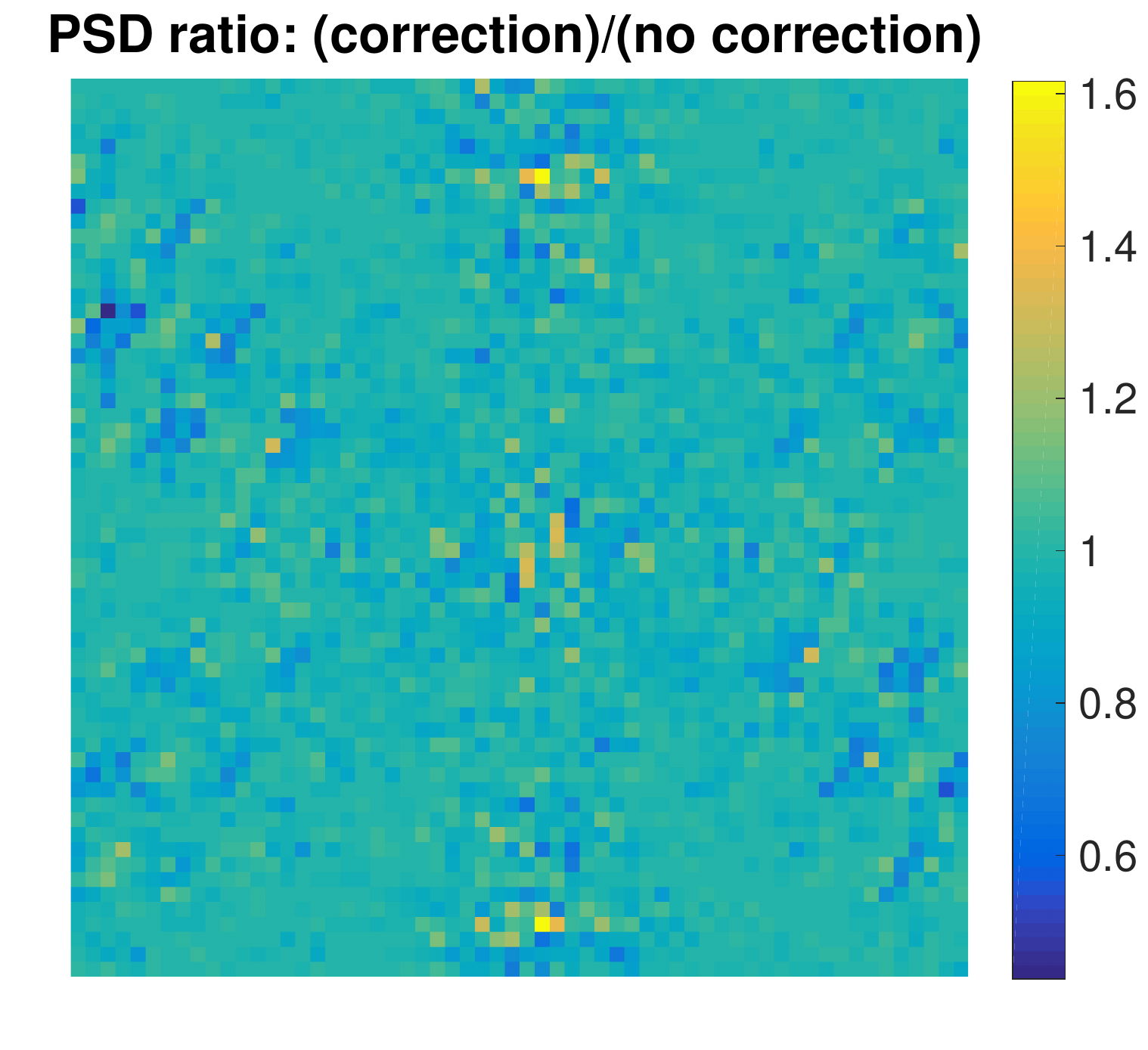}\includegraphics[width=0.5\columnwidth]{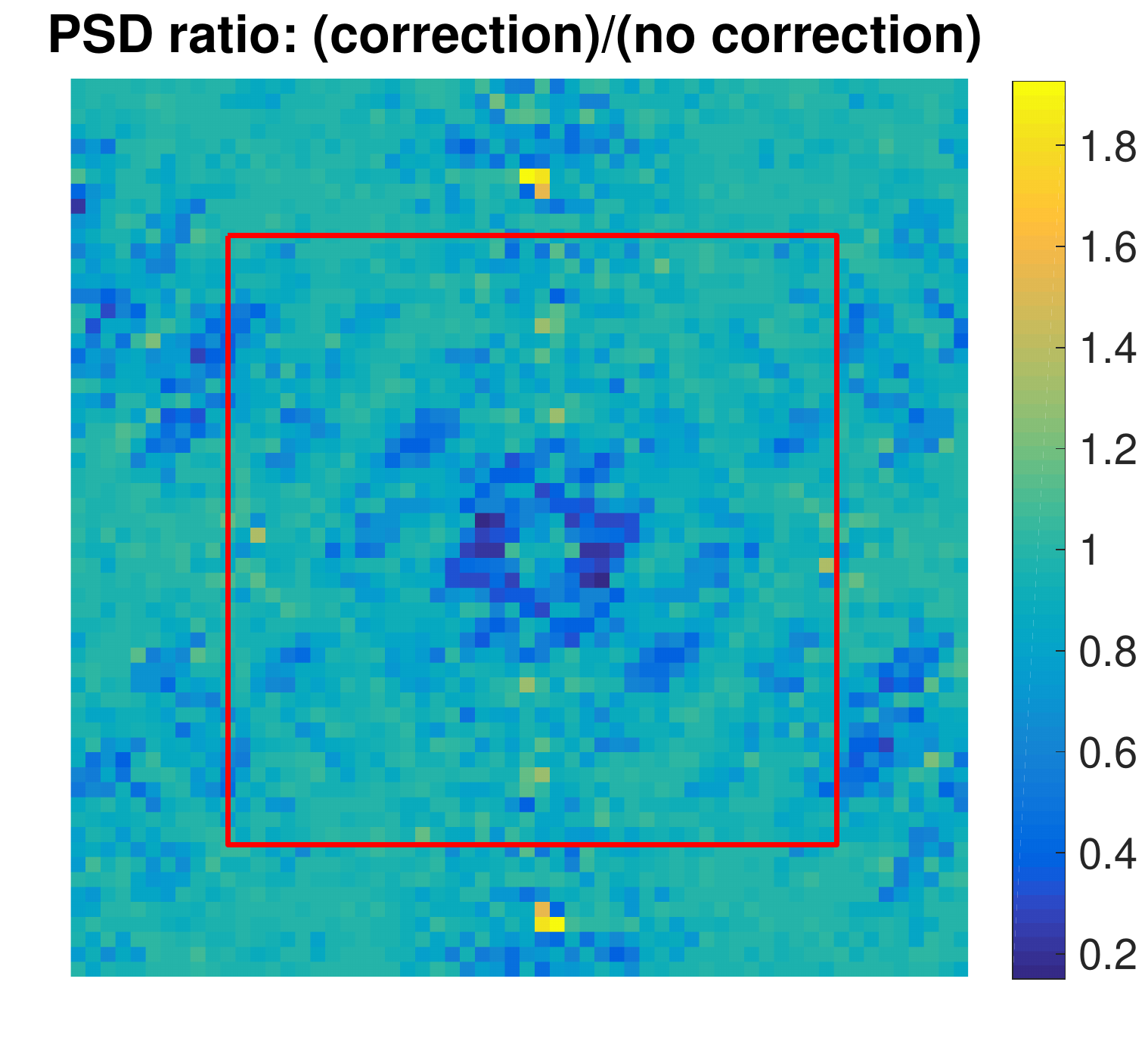}\\
 \caption{2D power spectral density ratios between uncorrected and corrected phasing errors. Left: \textit{conventional} phasing errors. Right: \textit{substantial} phasing errors. In the latter case, there is a clear improvement at lower spatial frequencies. Also shown as a red box is the control region of the DM, showing a clear `spillover' of high spatial frequencies outside the control radius due to the segment phasing errors.}
 \label{fig:psds}
\end{figure}

As expected, the resulting Strehl ratio improvement is significantly larger as indicated in Figure \ref{fig:strehls}, yielding a gain of 10\% at H-band. Equally striking is the improvement in regards to the PSD content, where it can be seen in Figure \ref{fig:psds} the major improvements at lower spatial frequencies when looking at the ratio between corrected and uncorrected 2D PSDs (shown in log stretch). For reference, the control region of the DM is shown as a red box; it is clear from this figure that `spillover' occurs beyond the control region when large phasing errors are present, impacting science applications such as high-contrast imaging, where controlling high spatial frequencies greatly improves contrast. The use of our algorithm corrects this `spillover'. The PSDs here are calculated from the average residual wavefront of the $\sim$1000 samples, gathered from 1 second of AO-telemetry. 

%% file: Sections/discussion.tex
We have demonstrated in simulation the effectiveness of using single, defocused images to simultaneously estimate primary mirror segment phasing errors and additional static errors in the form of NCPA. We first demonstrate the capabilities of this algorithm assuming the availability of an artificial source within the telescope structure that illuminates the primary; in this context the algorithm works extremely well, quantifying both phasing errors and NCPA to within a few percent. We note however that this approach does not take into account any local turbulence that may exist within the optical path of the source and furthermore - that such a source currently exists on such a telescope. However, any future segmented telescope with an AO-system could greatly benefit from such a source to perform focal plane wavefront error estimation.

The framework developed in the context of this artificial source scenario was expanded to consider simulated on-sky images of the Keck/NIRC2 system. We demonstrate that a set of 75 short exposure, defocused, AO-corrected images obtained within the timescale of one minute can successfully yield an estimation of both segment phasing errors and NCPA using our algorithm. Our algorithm reliably disentangles segmented phasing errors from NCPA, allowing one to consider applying a correction from such an estimate to the segmented primary and to the DM with confidence. Our approach relies on a slight modification to the current RTC (to employ both a WFS mask and extract full-turbulence slopes) and could potentially result in a total error reduction of 67\% and 57\% for realistic phasing errors and NCPA, respectively. Moreover, a direct application of our algorithm to the current Keck control scheme (i.e. with no modification to the RTC) could potentially result in an error reduction of 50\% and 65\% for realistic phasing errors and NCPA, respectively;  if the AO system applies a KL modal control with optimal mode filtration, then we can further mitigate the segment phasing error reduction to a total of 55\%. We note the significance of better NCPA estimation when considering the scenario with no modification to the RTC and advocate our algorithm could have significant potential applications in NCPA estimation alone for any telescope with an AO-system. Alternatively, a two-pronged approach to first estimate and correct the phasing errors reliably (using the RTC-altered method), followed up with an NCPA estimation and correction from a non-RTC altered scenario could yield optimal estimation of both types of aberrations (phasing errors and NCPA). A potential next step for our algorithm in the context of Keck/NIRC2 is to acquire a set of on-sky images that suit the requirements of our algorithm and to obtain an estimate of the phasing errors; this estimate could either be used as an immediate offset to the primary mirror segments and a direct assessment could be measured. Or (more conservatively) our estimate could be made just prior to an upcoming phasing run using the BBP technique and a form of comparison could be made between the two techniques.

In terms of alternative future work, current lab testing on the BESTFRENDS bench at the Dunlap Institute is aimed at confirming the findings of our simulation and preliminary results are promising (Lamb et al 2020, in prep.). Furthermore, there is a clear motivation to expand the study in the context of a pyramid WFS (PWFS) \citep{Ragazzoni1996} - particularly in the context of the near-infrared PWFS for Keck's KPIC \citep{Bond2018}. As such, we are currently establishing the framework of our algorithm to consider the application to a PWFS system such as KPIC. One particular aspect of exploration will be the NCPA quantification using on-sky images. Recent work has demonstrated reliable NCPA compensation with a pyramid via tracking of the optical gain \citep{Esposito2020}; this approach relies on the daytime quantification of NCPA using a calibration source. Using our algorithm would provide an estimation \textit{in situ}, providing the cumulative wavefront error estimate of the entire optical train down to the focal plane, free from any potential deviations from daytime measurements; a subsequent correction could immediately be made do the DM. In this context, our algorithm could be a potential improvement over existing on-line approaches, such as long-exposure Phase Diversity \citep{Mugnier2008}, where our algorithm is faster (requiring only short exposure images), potentially a more accurate alternative (given we can reliably isolate any NCPA from phasing errors), and relatively simple to implement. Given that the PWFS is the baseline for many future AO-instruments on GSMTs \citep[i.e.][]{Veran2015}, an exploration of our approach within this context is also clearly desirable.

%% file: main_final.bbl
\begin{thebibliography}{}
\makeatletter
\relax
\def\mn@urlcharsother{\let\do\@makeother \do\$\do\&\do\#\do\^\do\_\do\%\do\~}
\def\mn@doi{\begingroup\mn@urlcharsother \@ifnextchar [ {\mn@doi@}
  {\mn@doi@[]}}
\def\mn@doi@[#1]#2{\def\@tempa{#1}\ifx\@tempa\@empty \href
  {http://dx.doi.org/#2} {doi:#2}\else \href {http://dx.doi.org/#2} {#1}\fi
  \endgroup}
\def\mn@eprint#1#2{\mn@eprint@#1:#2::\@nil}
\def\mn@eprint@arXiv#1{\href {http://arxiv.org/abs/#1} {{\tt arXiv:#1}}}
\def\mn@eprint@dblp#1{\href {http://dblp.uni-trier.de/rec/bibtex/#1.xml}
  {dblp:#1}}
\def\mn@eprint@#1:#2:#3:#4\@nil{\def\@tempa {#1}\def\@tempb {#2}\def\@tempc
  {#3}\ifx \@tempc \@empty \let \@tempc \@tempb \let \@tempb \@tempa \fi \ifx
  \@tempb \@empty \def\@tempb {arXiv}\fi \@ifundefined
  {mn@eprint@\@tempb}{\@tempb:\@tempc}{\expandafter \expandafter \csname
  mn@eprint@\@tempb\endcsname \expandafter{\@tempc}}}

\bibitem[\protect\citeauthoryear{Bond et~al.,}{Bond et~al.}{2018}]{Bond2018}
Bond C.~Z.,  et~al., 2018, in Close L.~M.,  Schreiber L.,   Schmidt D.,  eds,
  Proc. SPIE Vol. 10703, Adaptive Optics Systems VI. SPIE, pp 642--652,
  \mn@doi{10.1117/12.2314121}

\bibitem[\protect\citeauthoryear{Chanan \& Troy}{Chanan \&
  Troy}{2018}]{Chanan2018}
Chanan G.,  Troy M.,  2018, in Marshall H.~K.,  Spyromilio J.,  eds,  Proc.
  SPIE Vol. 10700, Ground-based and Airborne Telescopes VII. SPIE, pp 415--425,
  \mn@doi{10.1117/12.2310157}, \url {https://doi.org/10.1117/12.2310157}

\bibitem[\protect\citeauthoryear{Chanan, Troy  \& Ohara}{Chanan
  et~al.}{2000}]{Chanan2000}
Chanan G.~A.,  Troy M.,   Ohara C.~M.,  2000, in Dierickx P.,  ed.,  Proc. SPIE
  Vol. 4003, Optical Design, Materials, Fabrication, and Maintenance. SPIE, pp
  188--202, \mn@doi{10.1117/12.391510}, \url
  {https://doi.org/10.1117/12.391510}

\bibitem[\protect\citeauthoryear{Conan \& Correia}{Conan \&
  Correia}{2014}]{OOMAO}
Conan R.,  Correia C.,  2014, in Adaptive Optics Systems IV. p. 91486C,
  \mn@doi{10.1117/12.2054470}

\bibitem[\protect\citeauthoryear{Dohlen, Wildi, Puget, Mouillet  \&
  Beuzit}{Dohlen et~al.}{2011}]{Dohlen2011}
Dohlen K.,  Wildi F.,  Puget P.,  Mouillet D.,   Beuzit J.-L.,  2011, in Second
  International Conference on Adaptive Optics for Extremely Large Telescopes.
  Online at {\textless}A
  href=''http://ao4elt2.lesia.obspm.fr''{\textgreater}http://ao4elt2.lesia.obspm.fr{\textless}/A{\textgreater},
  id.75.

\bibitem[\protect\citeauthoryear{Ellerbroek \& Vogel}{Ellerbroek \&
  Vogel}{2003}]{Ellerbroek2003}
Ellerbroek B.~L.,  Vogel C.~R.,  2003, in Tyson R.~K.,  Lloyd-Hart M.,  eds,
  Proc. SPIE Vol. 5169, Astronomical Adaptive Optics Systems and Applications.
  SPIE, pp 206--217, \mn@doi{10.1117/12.506580}, \url
  {https://doi.org/10.1117/12.506580}

\bibitem[\protect\citeauthoryear{Ellerbroek, Thelen, Lee, Carrara  \&
  Paxman}{Ellerbroek et~al.}{1997}]{Ellerbroek1997}
Ellerbroek B.~L.,  Thelen B.~J.,  Lee D.~J.,  Carrara D.~A.,   Paxman R.~G.,
  1997, in Tyson R.~K.,  Fugate R.~Q.,  eds,  Proc. SPIE Vol. 3126, Adaptive
  Optics and Applications. SPIE, pp 307--320, \mn@doi{10.1117/12.290157}, \url
  {https://doi.org/10.1117/12.290157}

\bibitem[\protect\citeauthoryear{Esposito et~al.,}{Esposito
  et~al.}{2010}]{Esposito2010}
Esposito S.,  et~al., 2010, \mn@doi [Appl. Opt.] {10.1364/AO.49.00G174}, 49,
  G174

\bibitem[\protect\citeauthoryear{Esposito, Puglisi, Pinna, Agapito,
  Quir{\'{o}}s-Pacheco, V{\'{e}}ran  \& Herriot}{Esposito
  et~al.}{2020}]{Esposito2020}
Esposito S.,  Puglisi A.,  Pinna E.,  Agapito G.,  Quir{\'{o}}s-Pacheco F.,
  V{\'{e}}ran J.~P.,   Herriot G.,  2020, \mn@doi [A{\&}A]
  {10.1051/0004-6361/201937033}, 636, A88

\bibitem[\protect\citeauthoryear{Gerard, Marois  \& Galicher}{Gerard
  et~al.}{2018}]{Gerard2018}
Gerard B.~L.,  Marois C.,   Galicher R.,  2018, \mn@doi [AJ]
  {10.3847/1538-3881/aad23e}, 156, 106

\bibitem[\protect\citeauthoryear{Gonsalves}{Gonsalves}{1982}]{Gonsalves82}
Gonsalves R.~A.,  1982, \mn@doi [Optical Engineering] {10.1117/12.7972989}, 21,
  215829

\bibitem[\protect\citeauthoryear{Guyon}{Guyon}{2004}]{Guyon2004}
Guyon O.,  2004, \mn@doi [ApJ] {10.1086/423980}, 615, 562

\bibitem[\protect\citeauthoryear{Hartung, Blanc, Fusco, Lacombe, Mugnier,
  Rousset  \& Lenzen}{Hartung et~al.}{2003}]{Hartung2003}
Hartung M.,  Blanc A.,  Fusco T.,  Lacombe F.,  Mugnier L.,  Rousset G.,
  Lenzen R.,  2003, \mn@doi [A{\&}A] {10.1051/0004-6361:20021744}, 399, 385

\bibitem[\protect\citeauthoryear{Heritier et~al.,}{Heritier
  et~al.}{2018}]{Heritier2018}
Heritier C.~T.,  et~al., 2018, \mn@doi [Monthly Notices of the Royal
  Astronomical Society] {10.1093/mnras/sty2485}, 481, 2829

\bibitem[\protect\citeauthoryear{Jolissaint, Mugnier, Neyman, Christou  \&
  Wizinowich}{Jolissaint et~al.}{2012}]{Jolissaint2012}
Jolissaint L.,  Mugnier L.,  Neyman C.,  Christou J.,   Wizinowich P.,  2012,
  in Adaptive Optics Systems III. p. 844716, \mn@doi{10.1117/12.926637}

\bibitem[\protect\citeauthoryear{Lamb et~al.,}{Lamb et~al.}{2017}]{Lamb2017b}
Lamb M.,  et~al., 2017, \mn@doi [Journal of Astronomical Telescopes,
  Instruments, and Systems] {10.1117/1.JATIS.3.3.039001}, 3, 39001

\bibitem[\protect\citeauthoryear{Lamb, Norton, {Macintosh B.}, {Correia C.},
  V{\'{e}}ran, {Marois C.}  \& Sivanandam}{Lamb et~al.}{2018}]{Lamb2018}
Lamb M.,  Norton M.,  {Macintosh B.} {Correia C.} V{\'{e}}ran J.-P.,  {Marois
  C.}  Sivanandam S.,  2018, in Close L.~M.,  Schreiber L.,   Schmidt D.,  eds,
   Proc. SPIE Vol. 10703, Adaptive Optics Systems VI. SPIE, pp 1344--1355,
  \mn@doi{10.1117/12.2313458}

\bibitem[\protect\citeauthoryear{Lofdahl, Kendrick, Harwit, Mitchell, Duncan,
  Seldin, Paxman  \& Acton}{Lofdahl et~al.}{1998}]{Lofdahl1998}
Lofdahl M.~G.,  Kendrick R.~L.,  Harwit A.,  Mitchell K.~E.,  Duncan A.~L.,
  Seldin J.~H.,  Paxman R.~G.,   Acton D.~S.,  1998, in Bely P.~Y.,
  Breckinridge J.~B.,  eds,  Proc. SPIE Vol. 3356, Space Telescopes and
  Instruments V. SPIE, pp 1190--1201, \mn@doi{10.1117/12.324519}, \url
  {https://doi.org/10.1117/12.324519}

\bibitem[\protect\citeauthoryear{Mahajan}{Mahajan}{1983}]{Mahajan1983}
Mahajan V.~N.,  1983, \mn@doi [J. Opt. Soc. Am.] {10.1364/JOSA.73.000860}, 73,
  860

\bibitem[\protect\citeauthoryear{Milli et~al.,}{Milli et~al.}{2018}]{Milli2018}
Milli J.,  et~al., 2018, in Close L.~M.,  Schreiber L.,   Schmidt D.,  eds,
  Proc. SPIE Vol. 10703, Adaptive Optics Systems VI. SPIE, pp 752--771,
  \mn@doi{10.1117/12.2311499}, \url {https://doi.org/10.1117/12.2311499}

\bibitem[\protect\citeauthoryear{Mugnier, Sauvage, Fusco, Cornia  \&
  Dandy}{Mugnier et~al.}{2008}]{Mugnier2008}
Mugnier L.~M.,  Sauvage J.-F.,  Fusco T.,  Cornia A.,   Dandy S.,  2008,
  \mn@doi [Opt. Express] {10.1364/OE.16.018406}, 16, 18406

\bibitem[\protect\citeauthoryear{N'Diaye, Martinache, Jovanovic, Lozi, Guyon,
  Norris, Ceau  \& Mary}{N'Diaye et~al.}{2018}]{N'Diaye2018}
N'Diaye M.,  Martinache F.,  Jovanovic N.,  Lozi J.,  Guyon O.,  Norris B.,
  Ceau A.,   Mary D.,  2018, \mn@doi [A{\&}A] {10.1051/0004-6361/201731985},
  610, A18

\bibitem[\protect\citeauthoryear{Paxman, Schulz  \& Fienup}{Paxman
  et~al.}{1992}]{Paxman92}
Paxman R.~G.,  Schulz T.~J.,   Fienup J.~R.,  1992, Journal of the Optical
  Society of America A, 9, 1027

\bibitem[\protect\citeauthoryear{Poyneer}{Poyneer}{2003}]{Poyneer2003}
Poyneer L.~A.,  2003, \mn@doi [Appl. Opt.] {10.1364/AO.42.005807}, 42, 5807

\bibitem[\protect\citeauthoryear{Ragazzoni}{Ragazzoni}{1996}]{Ragazzoni1996}
Ragazzoni R.,  1996, \mn@doi [Journal of Modern Optics]
  {10.1080/09500349608232742}, 43, 289

\bibitem[\protect\citeauthoryear{Ragland}{Ragland}{2018}]{Ragland2018}
Ragland S.,  2018, in Marshall H.~K.,  Spyromilio J.,  eds,  Proc. SPIE Vol.
  10700, Ground-based and Airborne Telescopes VII. SPIE, pp 407--414,
  \mn@doi{10.1117/12.2313017}, \url {https://doi.org/10.1117/12.2313017}

\bibitem[\protect\citeauthoryear{Ragland et~al.,}{Ragland
  et~al.}{2016}]{Ragland2016}
Ragland S.,  et~al., 2016, in Marchetti E.,  Close L.~M.,   V{\'{e}}ran J.-P.,
  eds,  Proc. SPIE Vol. 9909, Adaptive Optics Systems V. SPIE, pp 573--590,
  \mn@doi{10.1117/12.2231814}, \url {https://doi.org/10.1117/12.2231814}

\bibitem[\protect\citeauthoryear{Rampy, Ragland, Wizinowich  \& Campbell}{Rampy
  et~al.}{2014}]{Rampy2014}
Rampy R.,  Ragland S.,  Wizinowich P.,   Campbell R.,  2014, in Marchetti E.,
  Close L.~M.,   V{\'{e}}ran J.-P.,  eds,  Proc. SPIE Vol. 9148, Adaptive
  Optics Systems IV. SPIE, pp 1786--1802, \mn@doi{10.1117/12.2056902}, \url
  {https://doi.org/10.1117/12.2056902}

\bibitem[\protect\citeauthoryear{Roddier}{Roddier}{1981}]{Roddier1981}
Roddier F.,  1981, \mn@doi [Progess in Optics] {10.1016/S0079-6638(08)70204-X},
  19, 281

\bibitem[\protect\citeauthoryear{Sauvage, Mugnier, Paul  \& Villecroze}{Sauvage
  et~al.}{2012}]{Sauvage2012}
Sauvage J.~F.,  Mugnier L.,  Paul B.,   Villecroze R.,  2012, \mn@doi [Optics
  Letters] {10.1364/OL.37.004808}, 37, 4808

\bibitem[\protect\citeauthoryear{Sauvage et~al.,}{Sauvage
  et~al.}{2016}]{Sauvage2016}
Sauvage J.-F.,  et~al., 2016, in Marchetti E.,  Close L.~M.,   V{\'{e}}ran
  J.-P.,  eds,  Proc. SPIE Vol. 9909, Adaptive Optics Systems V. SPIE, pp
  408--416, \mn@doi{10.1117/12.2232459}, \url
  {https://doi.org/10.1117/12.2232459}

\bibitem[\protect\citeauthoryear{Travouillon, Els, Riddle, Sch{\"{o}}ck  \&
  Skidmore}{Travouillon et~al.}{2009}]{Travouillon2009}
Travouillon T.,  Els S.,  Riddle R.~L.,  Sch{\"{o}}ck M.,   Skidmore W.,  2009,
  \mn@doi [PASP] {10.1086/605295}, 121, 787

\bibitem[\protect\citeauthoryear{V{\'{e}}ran, Esposito, Spano, Herriot  \&
  Andersen}{V{\'{e}}ran et~al.}{2015}]{Veran2015}
V{\'{e}}ran J.-P.,  Esposito S.,  Spano P.,  Herriot G.,   Andersen D.,  2015,
  in AO4ELT4.

\bibitem[\protect\citeauthoryear{van Dam}{van Dam}{2017}]{vanDam2017}
van Dam M.,  2017, in AO4ELT5.

\bibitem[\protect\citeauthoryear{van Dam, Mignant  \& Macintosh}{van Dam
  et~al.}{2004}]{vanDam2004}
van Dam M.~A.,  Mignant D.~L.,   Macintosh B.~A.,  2004, \mn@doi [Appl. Opt.]
  {10.1364/AO.43.005458}, 43, 5458

\makeatother
\end{thebibliography}
